\documentclass[journal]{IEEEtran}

\usepackage{cite}
\usepackage{amsmath,amssymb,amsfonts}
\usepackage{algorithmic}
\usepackage{graphicx}
\usepackage{textcomp}
\def\BibTeX{{\rm B\kern-.05em{\sc i\kern-.025em b}\kern-.08em
    T\kern-.1667em\lower.7ex\hbox{E}\kern-.125emX}}

\usepackage{color}
\usepackage{bm}
\usepackage{algorithm}

\usepackage{hhline}

\ifCLASSOPTIONcompsoc
\usepackage[caption=false, font=normalsize, labelfont=sf, textfont=sf]{subfig}
\else
\usepackage[caption=false, font=footnotesize]{subfig}
\fi

\usepackage{soul}

\usepackage{babel}
\usepackage{threeparttable}

\begin{document}
\title{Efficient Near-Field Imaging Using Cylindrical MIMO Arrays}
\author{Shiyong~Li, ~\IEEEmembership{Member,~IEEE,}  
    Shuoguang~Wang,
    Moeness~G.~Amin, ~\IEEEmembership{Fellow,~IEEE,}
   and~Guoqiang~Zhao 
\thanks{Manuscript received 2020. 
	The work of Shiyong Li, Shuoguang Wang, and Guoqiang Zhao was supported by the National Natural Science Foundation of China under Grant 61771049.
     This work was initiated when Dr. Shiyong Li was a visiting scholar at the Center for Advanced Communications, Villanova University, PA, USA. }

\thanks{S. Li, S. Wang, and G. Zhao are with the Beijing Key Laboratory of Millimeter Wave and Terahertz Technology, Beijing Institute of Technology, Beijing 100081, China. (e-mail: lisy\_98@bit.edu.cn).}
\thanks{M. G. Amin is with the Center for Advanced Communications, Villanova University, Villanova, PA 19085 USA (e-mail: moeness.amin@villanova.edu).}

}
\markboth{IEEE}
{Shell \MakeLowercase{\textit{et al.}}: Wavenumber Domain Algorithm for Near-field Cylindrical MIMO Array Imaging}
\maketitle

\begin{abstract}
      
        
    Multiple-input multiple-output (MIMO) array based millimeter-wave (MMW) imaging has a tangible  prospect in applications of concealed weapons detection. A near-field imaging algorithm based on wavenumber domain processing is proposed for a cylindrical MIMO array scheme with uniformly spaced transmit and receive antennas over both the vertical and horizontal-arc directions. The spectrum aliasing associated with the proposed MIMO array is analyzed through a zero-filling discrete-time Fourier transform. The analysis shows that an undersampled array can be used in recovering the MMW image by a wavenumber domain algorithm. The requirements for the antenna inter-element spacing of the MIMO array are delineated. Numerical simulations as well as comparisons with the backprojection (BP) algorithm are provided to demonstrate the effectiveness of the proposed method.

\end{abstract}


\IEEEpeerreviewmaketitle

\section{Introduction}
Millimeter-wave (MMW) imaging can provide high target resolutions. Hence, it is applicable in broad sensing areas, such as remote sensing \cite{sar_bp,remote_sensing}, radio astronomy \cite{astronomy}, biomedical diagnosis \cite{biomedical}, indoor target tracking \cite{Amin_2010book,through_wall_amin}, and scattering diagnosis \cite{li1}.   
Since millimeter waves penetrate regular clothing, they can be used to form an accurate image of a person 
for detection of concealed objects. This type of imaging can be accomplished at moderate power levels without causing health hazards.  This attribute offers great potentials in personnel security inspections \cite{sheen}. Typically, a one-dimensional (1-D) antenna array with scanning  along only the perpendicular direction is used to gather the backscattered electromagnetic (EM) waves from the human body \cite{sheen}. 
Due to mechanical scanning, this approach has suffered  from slow data acquisition. The two-dimensional (2-D) antenna arrays with the full Nyquist samplings overcome this problem and meet the requirement of real-time imaging, but the system cost is often unaffordable.

The multiple-input multiple-output (MIMO)  systems have become ubiquitous and essential  for wireless communications \cite{mimo_commu}. MIMO system configuration has also been extensively applied in various radar applications \cite{mimo_rcs,mimo_radar2,mimo_radar3,mimo_hucheng,mimo_tian}. It  was recently utilized for near-field millimeter-wave imaging with reduced number of antenna elements required in a large aperture.  MIMO imaging systems enable gathering multistatic scattering information of the target which aids in mitigating the ghosts arising  with monostatic arrays \cite{ghosts_insight}. A high-resolution imaging system, combining the 1-D ultrawideband MIMO array and synthetic aperture radar (SAR), was proposed in \cite{zhuge1} for concealed weapons detection. This type of MIMO-SAR imaging scheme was also discussed in \cite{1dmimo_2, 1dmimo, 1d_mimo_cylindrical,guangyou,mimo_sar_wenqin} with different scanning apertures. 

Two-dimensional (2-D) MIMO arrays were examined in \cite{qps,zhuge_ap,zhuge2,2dmimo,tankai2,nufft_mimo2d} for near-field imaging. 
A 2-D planar MIMO system constructed by square clusters was developed in \cite{qps}, where back-projection (BP) algorithm was utilized for image reconstruction. 
In \cite{zhuge2}, the wavenumber domain algorithm, also named range migration, was proposed, in lieu of BP, for fast image reconstruction.
A version of this algorithm with improved image reconstructions, referred to as transverse spectrum deconvolution range migration, was considered in \cite{spectrum_deconvolution_2dmimo}.
A wavenumber domain algorithm based on a cross MIMO array was presented in \cite{cross_array_wda}. 
Parallelizable Fourier-based imaging algorithms were adopted for planar multistatic radar systems in \cite{wavenumber1, wavenumber2}.

Different from the aforementioned planar array topologies, we presented  a cylindrical MIMO array in \cite{cylindrical_mimo}. This array type  provides better observation angles than  planar arrays.  We employed an algorithm based on a multistatic-to-monostatic transformation along with an effective  phase calibration method \cite{planar_mimo}. 
However, approximation errors and imaging  restrictions remained. 
Therefore, for multistatic imaging, it has become important to deal directly with the scattered data. 

In this paper, we  propose an effective  wavenumber domain algorithm,  named cylindrical range migration algorithm for convenience,  for cylindrical MIMO array configurations. The transmit and receive antennas are uniformly spaced over both the vertical and horizontal-arc directions. 
To the best of our knowledge, MIMO arrays employing frequency-domain imaging algorithms have been only designed in accordance to the Nyquist sampling criterion \cite{1dmimo_2,zhuge2,1dmimo, 1d_mimo_cylindrical}. 
These designs reduce aperture exploitation efficiency, and offer reduced resolution relative to monostatic arrays with a same aperture size. 
In order to reach the same image resolution as a monostatic array, the transmit and receive MIMO arrays should have the same spatial frequency  extent (i.e., half extent of a monostatic array due to the one-way EM wave propagation related to either the transmit or the receive arrays). 
Then, in this situation and according to the Nyquist sampling criterion, the inter-element spacing of the transmit or receive array will be twice  that of a monostatic array. This means that the number of antennas of the MIMO configuration can only be reduced to a half of that of a monostatic array  which, in turn, limits and constrains MIMO array design.

Unlike existing works, the proposed  wavenumber-domain imaging technique, towards reducing the number of antennas, utilizes  undersampled subarrays associated with either the transmit or the receive arrays.
Both arrays occupy the same aperture size, as illustrated in Fig. \ref{cylindrical_mimo}. 
The spectrum aliasing in the underlying problem is analyzed based on the zero-filling discrete-time Fourier transform. 
It is shown that the proposed approach offers 
 imaging results similar to those of  time-domain algorithms, such as BP, implementing MIMO arrays, and similar resolutions as those of monostatic arrays. 

The rest of the paper is organized as follows. In Section II, we formulate the cylindrical range migration algorithm based on the spherical wave decompositions and the Fourier-domain convolutions. Several important issues, such as array dimension, spectrum analysis and processing, and sampling criteria are discussed in Section III. Numerical results are shown in Section IV, and concluding remarks follow at the end.

\section{Cylindrical MIMO Array Based Imaging}

\begin{figure}[!t]
	\centering
	\subfloat[]{\label{a}
		\includegraphics[width=2.6in]{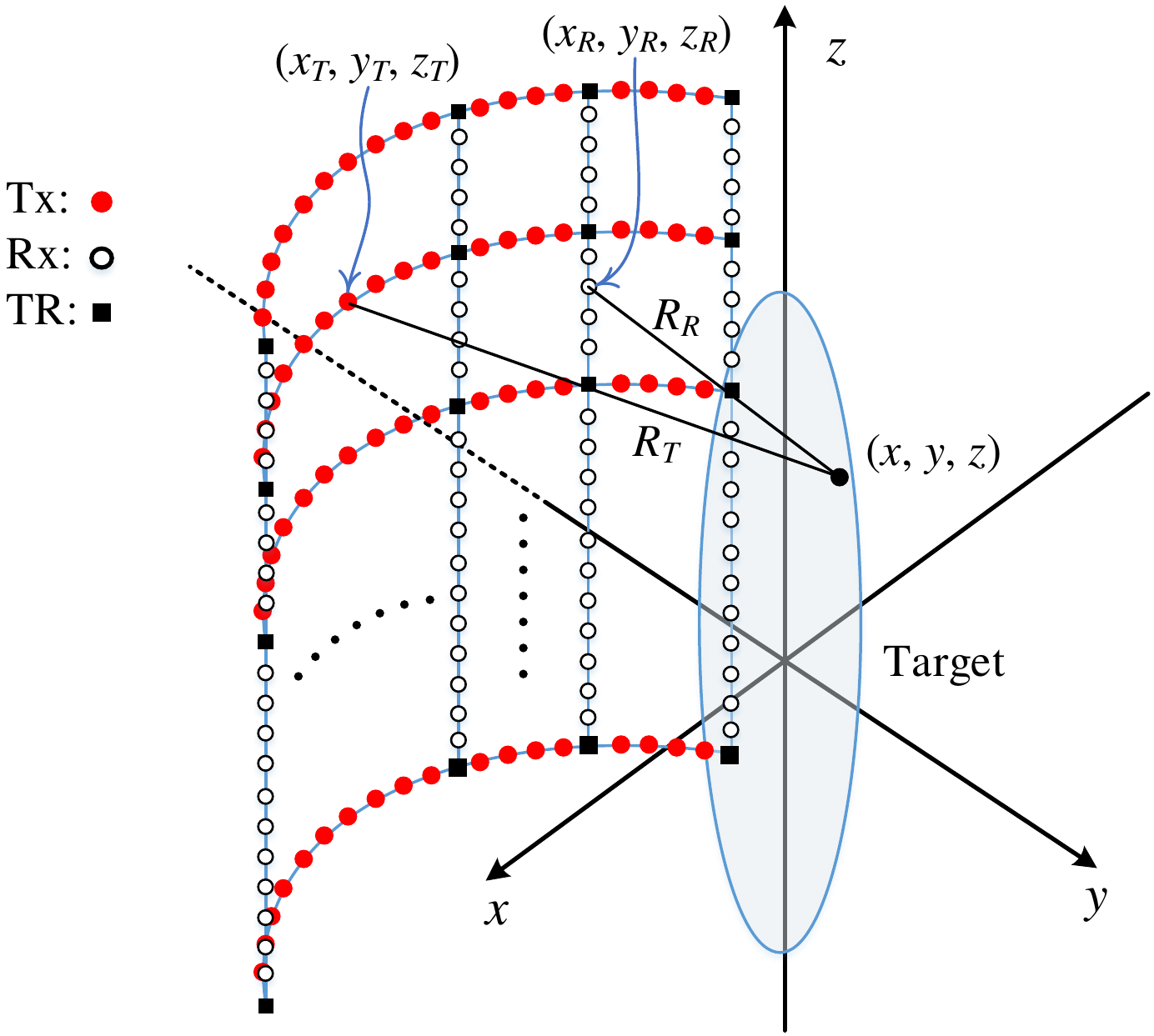}}
	\hfill
	\subfloat[]{\label{b}
		\includegraphics[width=1.7in]{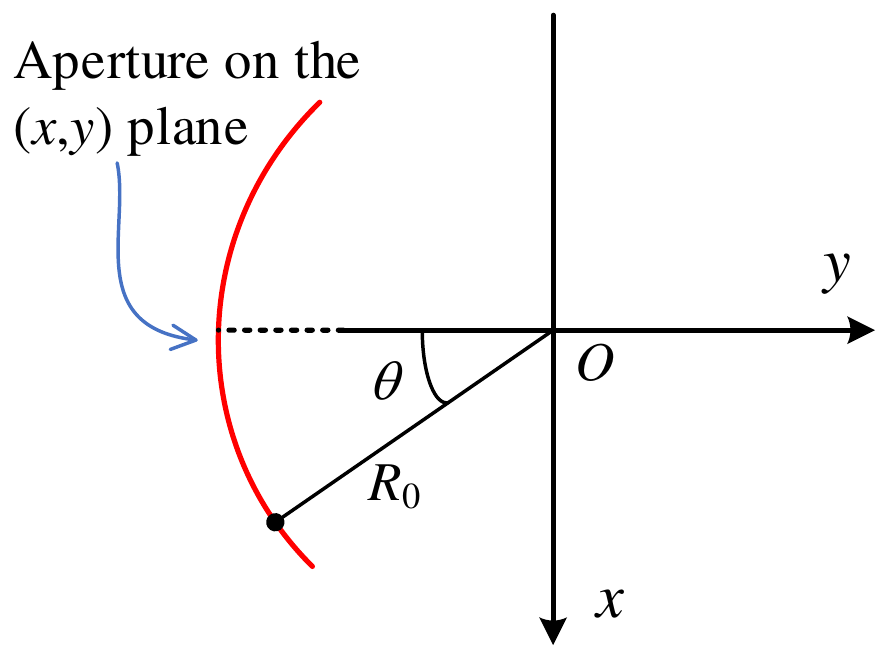}}
	\\
	
	\caption{(a) Topology of the cylindrical MIMO array, and (b) the geometrical relation on the $(x,y) $ plane.}
	\label{cylindrical_mimo}
\end{figure}

The MIMO imaging geometry is shown in Fig. \ref{cylindrical_mimo}. 
The transmit and receive antennas are uniformly spaced 
	on a cylindrical aperture. The transmit array meets the full sampling requirement along the horizontal arc direction, whereas it is uniform sparse along the vertical direction. The receive array assumes the opposite sampling configuration. 



The scattered EM waves from the target are given by, 
\begin{equation}\label{scat_wave1}
s(k,\theta_T,\theta_R,z_T,z_R)\!\!=\!\!\iiint \!\! g(x,y,z)e^{-\mathrm{j}k(R_T+R_R)}\mathrm{d}x\mathrm{d}y\mathrm{d}z, 
\end{equation}
where $k=\frac{2\pi f}{c}$ denotes the wavenumber, $f$ is the working frequency, $c$ is the speed of light, $g(x,y,z)$ represents the scattering coefficient of the target located at the Cartesian coordinate position $(x,y,z)$, $R_T$ and $R_R$ are, respectively, the distances from the transmit antenna to the target and from the target to the receive antenna, as shown in Fig. \ref{cylindrical_mimo}. These distances are  expressed by,
\begin{align}\nonumber
R_T &=\sqrt{\rho_T^2+(z-z_T)^2},  \\ 
R_R &=\sqrt{\rho_R^2+(z-z_R)^2},   \nonumber
\end{align}
where 
\begin{equation} \label{rhoT}
\rho_T=\sqrt{(x-x_T)^2+(y-y_T)^2},
\end{equation} 
and 
\begin{equation} \label{rhoR}
\rho_R=\sqrt{(x-x_R)^2+(y-y_R)^2}.
\end{equation}
The transmit and receive antenna Cartesian positions are denoted by  $(x_T,y_T,z_T)$ and $(x_R,y_R,z_R)$, respectively. They can also be represented in the cylindrical coordinates, according to $x_T=R_0\sin\theta_T$, $y_T=-R_0\cos\theta_T$, $x_R=R_0\sin\theta_R$, and  $y_R=-R_0\cos\theta_R$, where $R_0$ denotes the radius of the cylindrical array aperture, and $\theta$ is the angle between the negative direction of $y$-axis and the radius of the aperture in the $(x,y)$ plane, as depicted in Fig. \ref{cylindrical_mimo}\subref{b}.
Below, we present a wavenumber domain imaging algorithm using the cylindrical MIMO scheme. 

The exponential terms $e^{-\mathrm{j}kR_T}$ and $e^{-\mathrm{j}kR_R}$, appearing in \eqref{scat_wave1}, are referred to as the free space Green's functions, whose Fourier transforms with respect to $z_T$ and $z_R$ can be, respectively, expressed as \cite{soumekh}
\begin{equation} \label{grT}
\mathcal{F}_{z_T}[e^{-\mathrm{j}kR_T}]= e^{-\mathrm{j}(k_{\rho_T} \rho_T+k_{z_T}z)},
\end{equation}
\begin{equation} \label{grR}
\mathcal{F}_{z_R}[e^{-\mathrm{j}kR_R}]= e^{-\mathrm{j}(k_{\rho_R} \rho_R+k_{z_R}z)},
\end{equation}
where
\begin{subequations}
	\begin{align} 
	&k_{\rho_T}=\sqrt{k^2-k^2_{z_T}},\label{krhot} \\ 
    &k_{\rho_R}=\sqrt{k^2-k^2_{z_R}}.\label{krhor}
	\end{align}
\end{subequations}

Using the above equations, we proceed to apply the Fourier transforms to both sides of \eqref{scat_wave1} with respect to $z_T$ and $z_R$, 
\begin{align} \label{scat_wave3}
\widetilde{s}(k,\theta_T,\theta_R,&k_{z_T},k_{z_R})\!=\!\iiint g(x,y,z)\cdot\\ \nonumber
&e^{-\mathrm{j}(k_{\rho_T} \rho_T+k_{z_T}z)} e^{-\mathrm{j}(k_{\rho_R} \rho_R+k_{z_R}z)}\mathrm{d}x\mathrm{d}y\mathrm{d}z.
\end{align}
Based on \eqref{rhoT} and \eqref{rhoR}, and using the inverse Fourier transform expression,  
we decompose the cylindrical wave associated with $(R_0\sin\theta, -R_0\cos\theta)$, for $\theta=\theta_T$ or $\theta_R$, into a superposition of plane wave components \cite{soumekh},
\begin{align} \label{cylindrical2planeT}
e^{-\mathrm{j}k_{\rho_T} \rho_T}=\int e^{-\mathrm{j}k_{x_T}(x-x_T)}e^{-\mathrm{j}k_{y_T}(y-y_T)}\mathrm{d}k_{x_T},
\end{align}
\begin{align} \label{cylindrical2planeR}
e^{-\mathrm{j}k_{\rho_R} \rho_R}=\int e^{-\mathrm{j}k_{x_R}(x-x_R)}e^{-\mathrm{j}k_{y_R}(y-y_R)}\mathrm{d}k_{x_R},
\end{align}
where $k_{\rho_T}=\sqrt{k_{x_T}^2+k_{y_T}^2}$, and $k_{\rho_R}=\sqrt{k_{x_R}^2+k_{y_R}^2}$. 

Substituting \eqref{cylindrical2planeT} and \eqref{cylindrical2planeR} in \eqref{scat_wave3}, and rearranging the integral sequence, we obtain,
\begin{align} \label{scat_wave4}
\widetilde{s}(k,\theta_T,&\theta_R,k_{z_T},k_{z_R})\!=\!\iint\!\!\iiint g(x,y,z)\cdot\\ \nonumber
&e^{-\mathrm{j}(k_{x_T}+k_{x_R})x}e^{-\mathrm{j}(k_{y_T}+k_{y_R})y}e^{-\mathrm{j}(k_{z_T}+k_{z_R})z}\cdot\\ \nonumber
&e^{\mathrm{j}k_{x_T}x_T}e^{\mathrm{j}k_{y_T}y_T}e^{\mathrm{j}k_{x_R}x_R}e^{\mathrm{j}k_{y_R}y_R}\mathrm{d}x\mathrm{d}y\mathrm{d}z\mathrm{d}k_{x_T}\mathrm{d}k_{x_R}. 
\end{align}
The inner integrals over $x$, $y$, and $z$ can be expressed as a three-dimensional Fourier transform of $g$, denoted by $G(k_x,k_y,k_z)$,
\begin{align} \label{scat_wave5}
\widetilde{s}(k,\theta_T,&\theta_R,k_{z_T},k_{z_R})\!=\!\iint G(k_x,k_y,k_z)\cdot\\ \nonumber
&e^{\mathrm{j}k_{x_T}x_T}e^{\mathrm{j}k_{y_T}y_T}e^{\mathrm{j}k_{x_R}x_R}e^{\mathrm{j}k_{y_R}y_R}\mathrm{d}k_{x_T}\mathrm{d}k_{x_R}, 
\end{align}
where 
\begin{subequations}
	\begin{align} 
	&k_x=k_{x_T}+k_{x_R}, \label{kx} \\ 
	&k_y=k_{y_T}+k_{y_R}, \label{ky} \\ 
	&k_z=k_{z_T}+k_{z_R}, \label{kz}
	\end{align}
\end{subequations}


The relation of differential elements between the Cartesian and the polar coordinates is $\mathrm{d}k_{x_T}\mathrm{d}k_{y_T}=k_{\rho_T}\mathrm{d}k_{\rho_T}\mathrm{d}\phi_T$, where $\phi_T$ is the angle between $k_{\rho_T}$-axis and $-k_y$-axis, satisfying $k_{x_T}=k_{\rho_T}\sin\phi_T$ and $k_{y_T}=-k_{\rho_T}\cos\phi_T$.  
We maintain that  $\mathrm{d}k_{y_T}\approx -\mathrm{d}k_{\rho_T}$ for small values of  $\phi_T$. Accordingly, $\mathrm{d}k_{x_T} \approx -k_{\rho_T}\mathrm{d}\phi_T$ (So is the receive part).
Representing the space wavenumbers in the cylindrical coordinates, we obtain
\begin{align} \label{scat_wave7}
\widetilde{s}(k,&\theta_T,\theta_R,k_{z_T},k_{z_R})\!=\!k_{\rho_T}k_{\rho_R}\cdot \\ \nonumber
\!\iint \! &G(k_{\rho_T},k_{\rho_R},\phi_T,\phi_R,k_z)\cdot\\ \nonumber
&e^{\mathrm{j}k_{\rho_T}R_0\cos(\theta_T-\phi_T)}e^{\mathrm{j}k_{\rho_R}R_0\cos(\theta_R-\phi_R)}\mathrm{d}\phi_{T}\mathrm{d}\phi_{R}, 
\end{align}
where the wavenumber relations between the polar and Cartesian coordinates are given by,
\begin{subequations}
	\begin{align} 
	&k_{x_{T/R}}=k_{\rho_{T/R}}\sin\phi_{T/R}, \label{kxtr}\\
	&k_{y_{T/R}}=-k_{\rho_{T/R}}\cos\phi_{T/R}, \label{kytr} 
	\end{align}
\end{subequations}
with the subscript $``T/R"$ meaning the transmit or the receive part.

The above integrals over $\phi_{T}$ and $\phi_{R}$ can be represented by the convolutions with respect to $\theta_T$ and $\theta_R$, respectively.
\begin{align} \label{scat_wave8}
\widetilde{s}(k,\theta_T,&\theta_R,k_{z_T},k_{z_R})\!=\!k_{\rho_T}k_{\rho_R}G(k_{\rho_T},k_{\rho_R},\theta_T,\theta_R,k_z)\circledast_T \\ \nonumber
&e^{\mathrm{j}k_{\rho_T}R_0\cos\theta_T}\circledast_R e^{\mathrm{j}k_{\rho_R}R_0\cos\theta_R}, 
\end{align}
where $\circledast_T$ and $\circledast_R$ denote  convolutions in the  $\theta_T$ and $\theta_R$ domains, respectively. 

The imaging involves the deconvolutions of the two exponential functions, which are performed by applying  the Fourier transforms to both sides of \eqref{scat_wave8} with respect to $\theta_T$ and $\theta_R$. Using the Fourier convolution property, we can write
\begin{align} \label{scat_wave9}
\widetilde{\widetilde{s}}(k,\xi_T,&\xi_R,k_{z_T},k_{z_R})\!=\!k_{\rho_T}k_{\rho_R} \widetilde{G}(k_{\rho_T},k_{\rho_R},\xi_T,\xi_R,k_z)\cdot \\ \nonumber
&H^{(1)}_{\xi_T}(k_{\rho_T}R_0)e^{\mathrm{j}\pi\xi_T/2}H^{(1)}_{\xi_R}(k_{\rho_R} R_0)e^{\mathrm{j}\pi\xi_R/2}, 
\end{align}
where $\xi_{T/R}$ denotes the Fourier domain for $\theta_{T/R}$, and $H^{(1)}_{\xi_{T/R}}$ is the Hankel function of the first kind, $\xi_{T/R}$ order. 
\begin{equation}\nonumber
	H^{(1)}_{\xi_{T/R}}(k_{\rho_{T/R}} R_0)=\mathcal{F}_{\theta_{T/R}}[e^{\mathrm{j}k_{\rho_{T/R}}R_0\cos\theta_{T/R}}]e^{-\mathrm{j}\pi\xi_{T/R}/2}.
\end{equation}
The Hankel function can be determined analytically for $\xi\ll k_{\rho}R_0$ \cite{soumekh,sheen_cyindrical} as follows,
\begin{equation}\label{hankel2}
H^{(1)}_{\xi}(k_{\rho} R_0)=e^{\mathrm{j}\sqrt{k_\rho^2 R_0^2-\xi^2}}e^{-\mathrm{j}\pi\xi/2}.
\end{equation}

By dividing both sides of \eqref{scat_wave9} by the two Hankel functions and exponentials, and performing the inverse Fouirer transforms with respect to $\xi_{T}$ and $\xi_{R}$, we  obtain
\begin{align} \label{freq_target}
&G(k_{\rho_T},k_{\rho_R},\theta_T,\theta_R,k_{z_T},k_{z_R})=\\\nonumber
&\mathcal{F}^{-1}_{\xi_{T/R}}\Bigg[{\frac{\widetilde{\widetilde{s}}(k,\xi_T,\xi_R,k_{z_T},k_{z_R})e^{-\mathrm{j}\pi\xi_T/2}e^{-\mathrm{j}\pi\xi_R/2}}{k_{\rho_T}k_{\rho_R}H^{(1)}_{\xi_T}(k_{\rho_T}R_0)H^{(1)}_{\xi_R}(k_{\rho_R} R_0)}}\Bigg].
\end{align}
The left side of the above equation expresses dependence on $k_{\rho_T}$ and $k_{\rho_R}$ as these variables relate to $k$ according to \eqref{krhot} and \eqref{krhor}.
Finally, interpolations and dimension reduction are needed to bring $G(k,\theta_T,\theta_R,k_{z_T},k_{z_R})$ into the form of $G(k_x,k_y,k_z)$, over which the 3-D inverse fast Fourier transform (IFFT) is performed to yield $g(x,y,z)$. The details will be discussed in the next section.

\section{Several Important Issues}
\subsection{Interpolations and Dimension Change}


To properly perform interpolations from the polar coordinates $(k_{\rho_{T/R}},\theta_{T/R})$ to the Cartesian  $(k_{x_{T/R}},k_{y_{T/R}})$, we should change $G(k,\theta_T,\theta_R,k_{z_T},k_{z_R})$ to $G(k_T,\theta_T,k_R,\theta_R,k_{z_T},k_{z_R})$ via higher dimension manipulation using   
\begin{equation}
	k=k_T+k_R
\end{equation}
where 
\begin{subequations}
	\begin{align} 
	&k_T=\frac{1}{2}\sqrt{k^2_{\rho_T}+k^2_{z_T}},\label{kt} \\ 
	&k_R=\frac{1}{2}\sqrt{k^2_{\rho_R}+k^2_{z_R}},\label{kr}
	\end{align}
\end{subequations}
which are based on \eqref{krhot} and \eqref{krhor}. 

\begin{figure}[!t]
	\centering
	\includegraphics[width=2.5in]{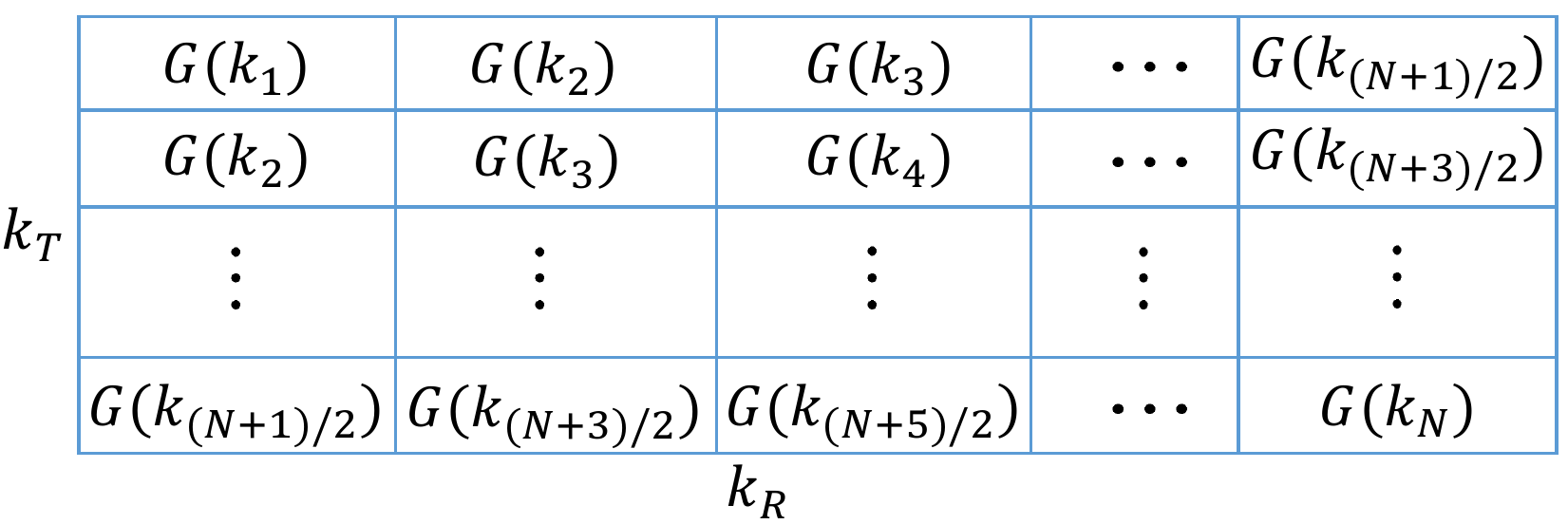}
	\caption{Illustration of the dimension increasing from $G(k,\cdots)$ to $G(k_T,k_R,\cdots)$.}
	\label{dim_increa}
\end{figure}

We then reformulate the function $G(k,\cdots)$  based on the relationship between the variables depicted in Fig. \ref{dim_increa}.
This provides the data matrix corresponding to the independent spectrum support $(k_T,\theta_T)$ and $(k_R,\theta_R)$, respectively.
$G(k_{\rho_T},\theta_T,k_{\rho_R},\theta_R,k_{z_T},k_{z_R})$ can then be obtained by employing \eqref{kt} and \eqref{kr}.
Next, we acquire  $G(k_{x_T},k_{y_T},k_{x_R},k_{y_R},k_{z_T},k_{z_R})$ via interpolations of the data in $(k_{\rho_{T/R}},\theta_{T/R})$ to $(k_{x_{T/R}},k_{y_{T/R}})$ in accordance with  \eqref{kxtr} and \eqref{kytr}. 

Finally, dimension reduction is applied to obtain $G(k_x,k_y,k_z)$ based on \eqref{kx}, \eqref{ky}, and \eqref{kz}, using the fact that the ideal sub-matrices related to $(k_{x_T},k_{x_R})$, $(k_{y_T},k_{y_R})$, and $(k_{z_T},k_{z_R})$ are symmetric. The symmetry property emerges if we choose the grids of the spatial frequencies of the transmit and receive arrays to be the same. 
This first requires the array inter-element spacing  along the vertical direction to satisfy the equality \cite{zhuge2},

\begin{align}
&\frac{1}{N_{z_T}\Delta z_T}=\frac{1}{N_{z_R}\Delta z_R}
\end{align}
where $\Delta z_T$ and $\Delta z_R$ are the antenna  spacings of the transmit and receive  arrays, respectively. $N_{z_T}$ and $N_{z_R}$ denote the points of fast Fourier transforms (FFTs) which transform the data from $(z_T,z_R)$ domain to $(k_{z_T},k_{z_R})$ domain.


Since the data in $(k_{x_T},k_{y_T},k_{x_R},k_{y_R})$ is generated by interpolations, then we can readily set the grids to be the same for the transmitter and receiver. 
However, we should carefully perform FFTs with respect to $\theta_T$ and $\theta_R$ to ensure the accuracy of convolutions in the frequency domain.

\begin{figure}[!t]
	\centering
	\includegraphics[width=3.2in]{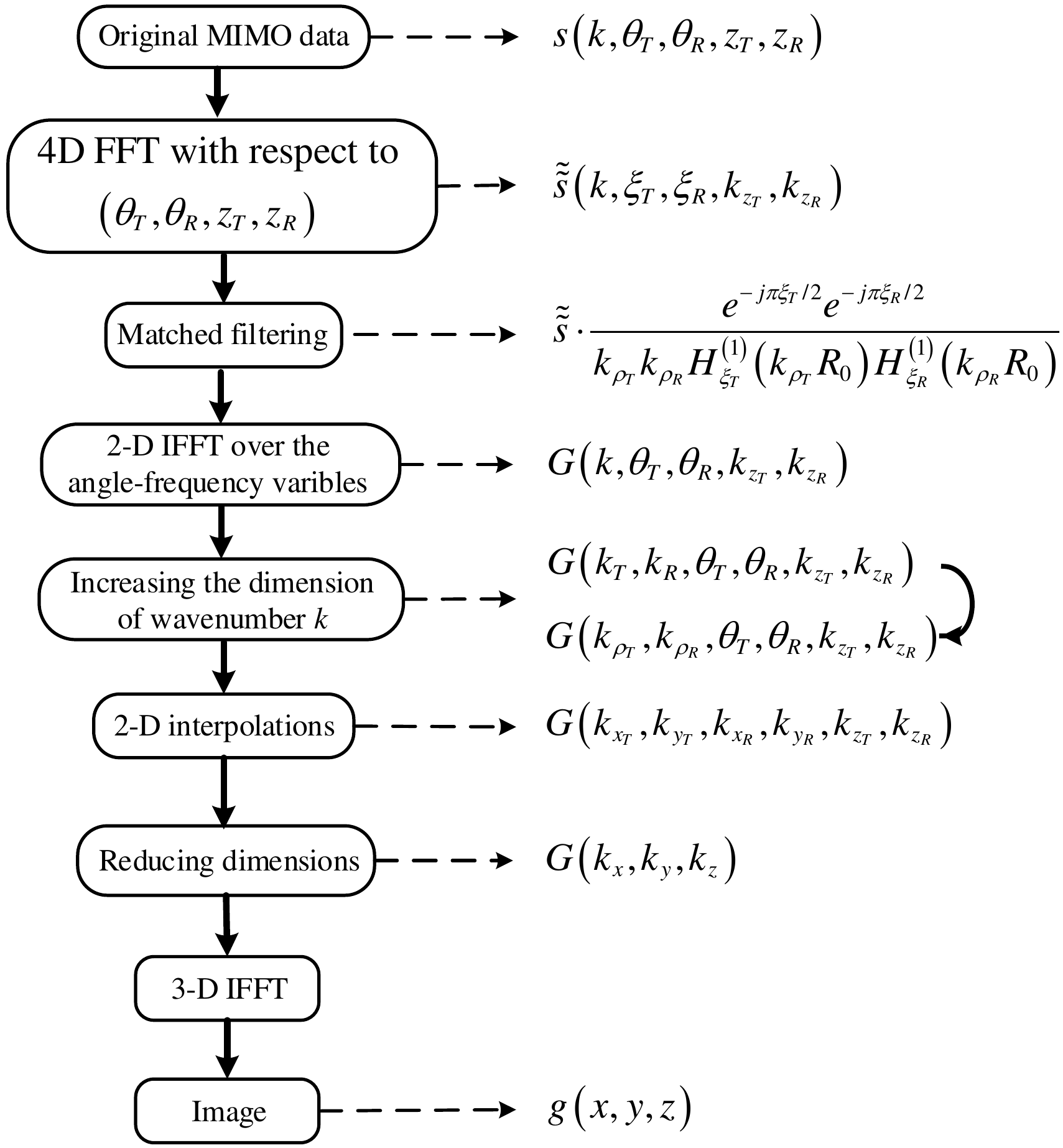}
	\caption{Block diagram of the  cylindrical RMA.}
	\label{cRMA}
\end{figure}

The block diagram of the complete imaging procedure  considered is shown in Fig. \ref{cRMA}.





\subsection{Spectrum Aliasing} 

Existing work based on  frequency-domain processing has only considered the case in which both the transmit and receive arrays satisfy the Nyquist sampling criteria \cite{1dmimo_2, zhuge2, 1dmimo, 1d_mimo_cylindrical}. However, in this case, either the number of antenna elements cannot be significantly reduced, or the image resolution is worse than the monostatic counterpart with a same aperture size.  

\subsubsection{Spectrum Aliasing Based on the Discrete-time Fourier Transform}

To fully utilize the aperture, we configure the transmit and receive arrays as shown in Fig. \ref{cylindrical_mimo}. 
Consider first the following discrete-time Fourier transform (DTFT) of temporal-domain data $s_P$ obtained by placing $P-1$ zeros between successive values of $s$ \cite{2015signals},
\begin{align} \label{fft_filling}
	S_P(e^{\mathrm{j}\omega})&=\sum_{n=-\infty}^{\infty} s_P(nP)e^{-\mathrm{j}\omega nP} \\ \nonumber
	&=\sum_{n=-\infty}^{\infty} s(n)e^{-\mathrm{j}P\omega n}=S(e^{\mathrm{j}P\omega})
\end{align}
where 
$n$ denotes discrete time 
, and $\omega$ repsents the frequency satisfying $\omega = 2\pi f$. 
Note that the corresponding Fourier transform $S_P(e^{\mathrm{j}\omega})$ is compressed to have a new period $2\pi/P$ in comparison to $S(e^{\mathrm{j}\omega})$ \cite{2015signals}. 
In other words, there are more periods of  spectrum of the original data within the $2\pi$ segment. 

\begin{figure}[!t]
	\centering
	\subfloat[]{\label{a}
		\includegraphics[width=2.3in]{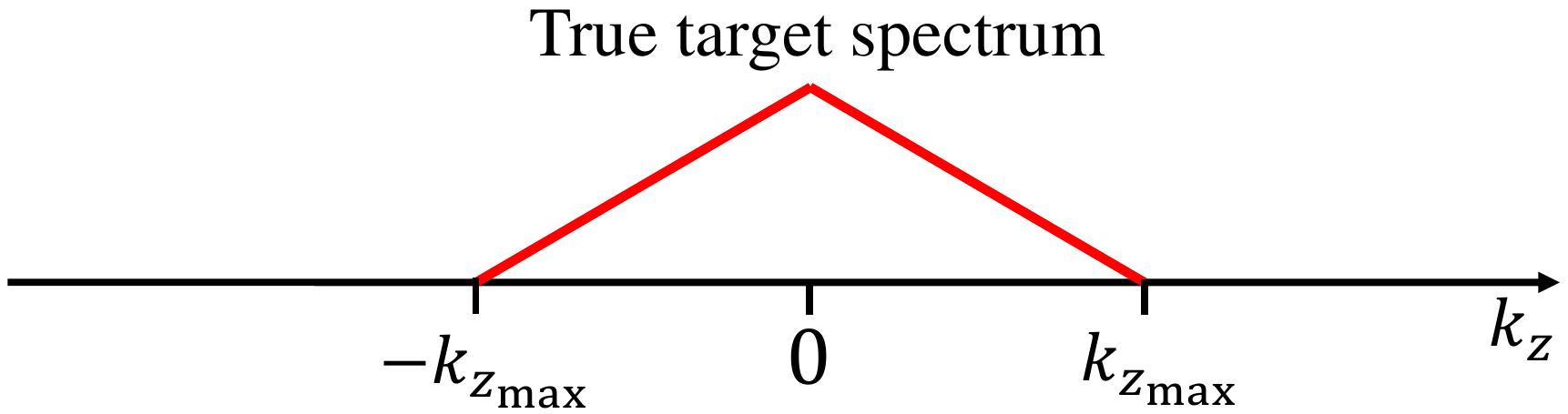}}
	\hfill
	\subfloat[]{\label{b}
		\includegraphics[width=2.3in]{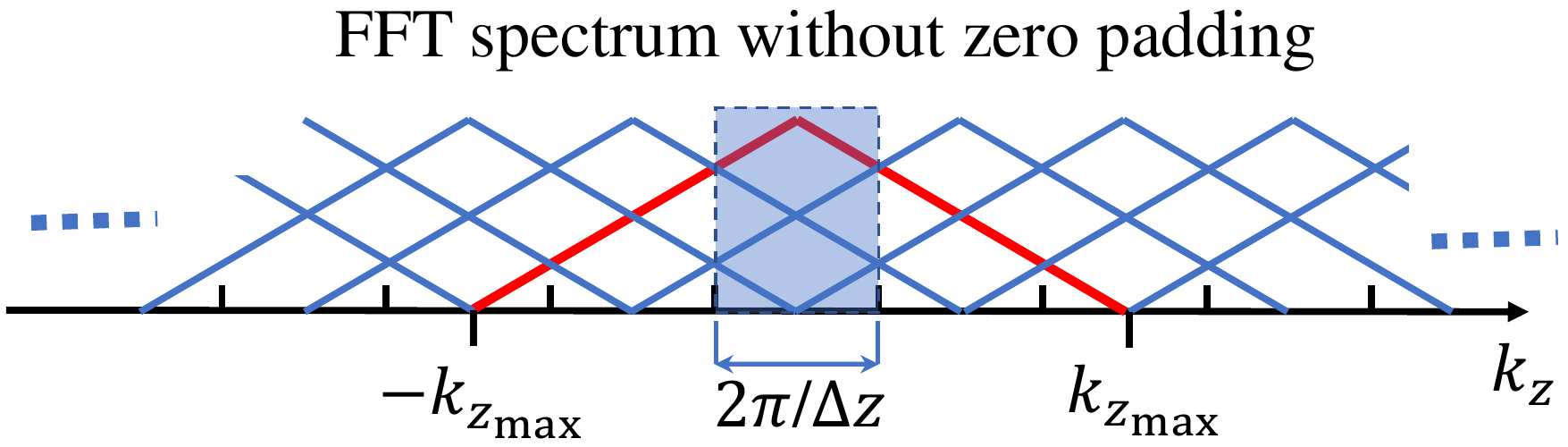}}
	\hfill
	\subfloat[]{\label{c}
		\includegraphics[width=2.5in]{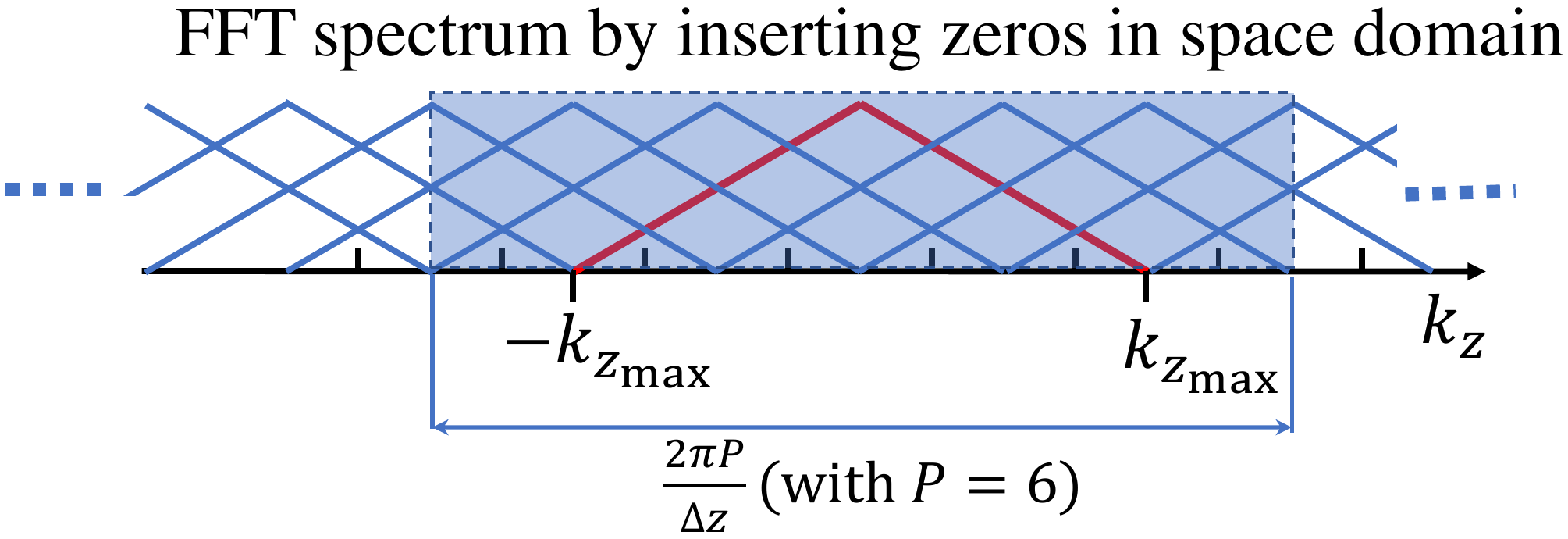}}
	\\
	\caption{(a) The spectrum over $k_z$, (b) the spectrum (shadow part) obatined by FFT for an undersampling scenario, and (c) the spectrum (shadow part) obtained by zero-filling FFT for the same case (assuming that ($P-1$) zeros are placed between successive values of the data in space domain).}
	\label{spectrum_analysis}
\end{figure}

\begin{figure}[!t]
	\centering
	\subfloat[]{\label{a}
		\includegraphics[width=1.3in]{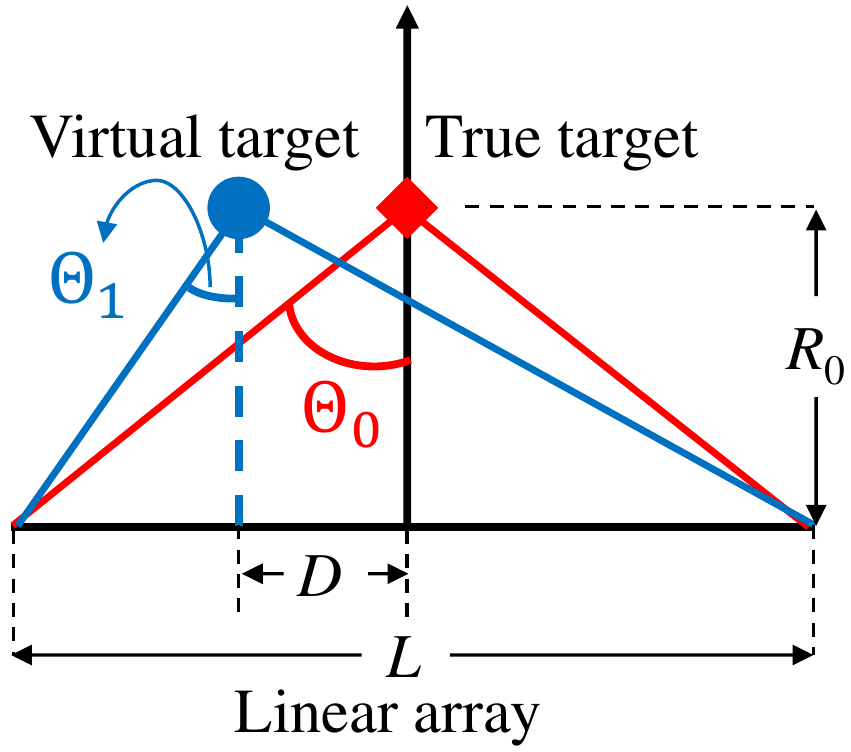}}
	\hfill
	\subfloat[]{\label{b}
		\includegraphics[width=1.3in]{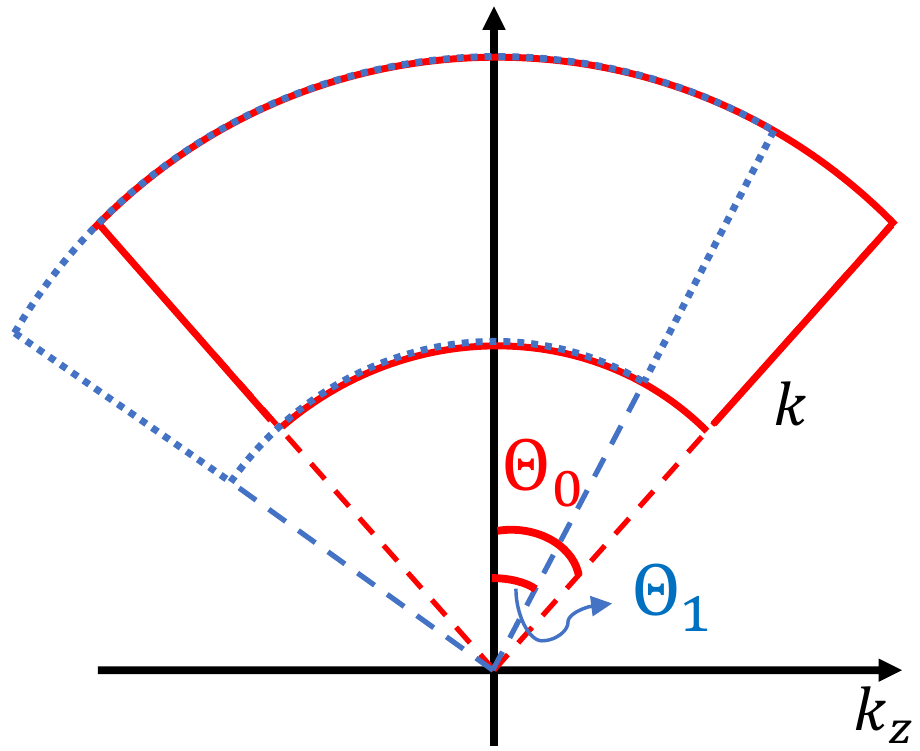}}
	\\
	
	\caption{Targets and their corresponding spectra, (a) two targets located at different positions (with a same down-range), and (b) the corresponding spectra.}
	\label{targets_spectra}
\end{figure}

The above relationship can be extended to the spatial frequency domain. 
Consider the simple example of a linear sparse monostatic array along the $z$ direction, with the inter-element spacing $\Delta z$ that is larger than the Nyquist sample spacing. Assume one point target is located at a distance $R_0$ away from the array center.      
The data in the spatial frequency domain obtained by the Fourier transforms of the received echo across the array are depicted in Fig. \ref{spectrum_analysis}. 
The true target wavenumber domain spectrum is shown in Fig. \ref{spectrum_analysis}\subref{a} which is determined by the geometrical relation between the target and array. 
Here, we use triangles for  illustration of different periods of the spectrum.  
If we  perform FFT of the data across  the sparse array without zero padding, the obtained  spectrum is indicated by the shadow part of the overlapping replicas shown in  Fig. \ref{spectrum_analysis}\subref{b}.  
On the other hand, the spectrum associated with zero padded data 
is illustrated in Fig. \ref{spectrum_analysis}\subref{c} (the shadow part).
Although spectrum aliasing is still present, the complete spectrum distribution exists within the shadow part, since the FFT extent is enlarged to $2\pi P/\Delta z$.  
The other parts can be viewed as spectra corresponding to targets  located at different positions relative to the true one. This is depicted in Fig. \ref{targets_spectra}, where only one aliasing target is included.

If we filter the spectrum according to the extent of the true target spectrum, i.e., $2k_{z_\text{max}}$, as depicted in Fig. \ref{spectrum_analysis}\subref{a}, then the result is more like that of tomographic processing. 
From the perspective of tomography, the demodulated echo signal can be interpreted as the approximate Fourier transform of the projection of the target to be imaged \cite{cbp_sar}. 
Accordingly, we can acheive a similar imaging result with that of BP by using a frequency-domain algorithm. 


\subsubsection{Example of a 1-D MIMO Array}

\begin{figure}[!t]
	\centering
	\includegraphics[width=1.9in]{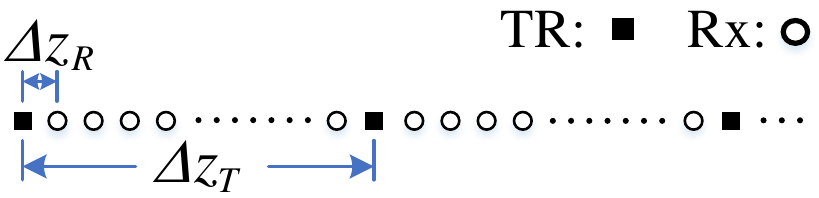}
	\caption{Illustration of a 1-D MIMO array.}
	\label{1dmimo}
\end{figure}

We consider the scenario of a linear MIMO array along the $z$ direction, consisting of a  receive array satisfying the Nyquist sampling criterion (with the inter-element spacing $\Delta z_R$) and a uniform sparse transmit array (with the inter-element spacing $\Delta z_T=P\Delta z_R$), both having a same length, as shown in Fig. \ref{1dmimo}. 
We illustrate the aforementioned aliasing effect by evaluating the performance of the near-field beam patterns \cite{beam_pattern} of the transmit and the  receive arrays, respectively. Here, we choose the array length as $L=1$m, the working frequency as $f=30$GHz, and set $P=20$. 


\begin{figure}[!t]
	\centering
	\subfloat[]{\label{a}
		\includegraphics[width=1.69in]{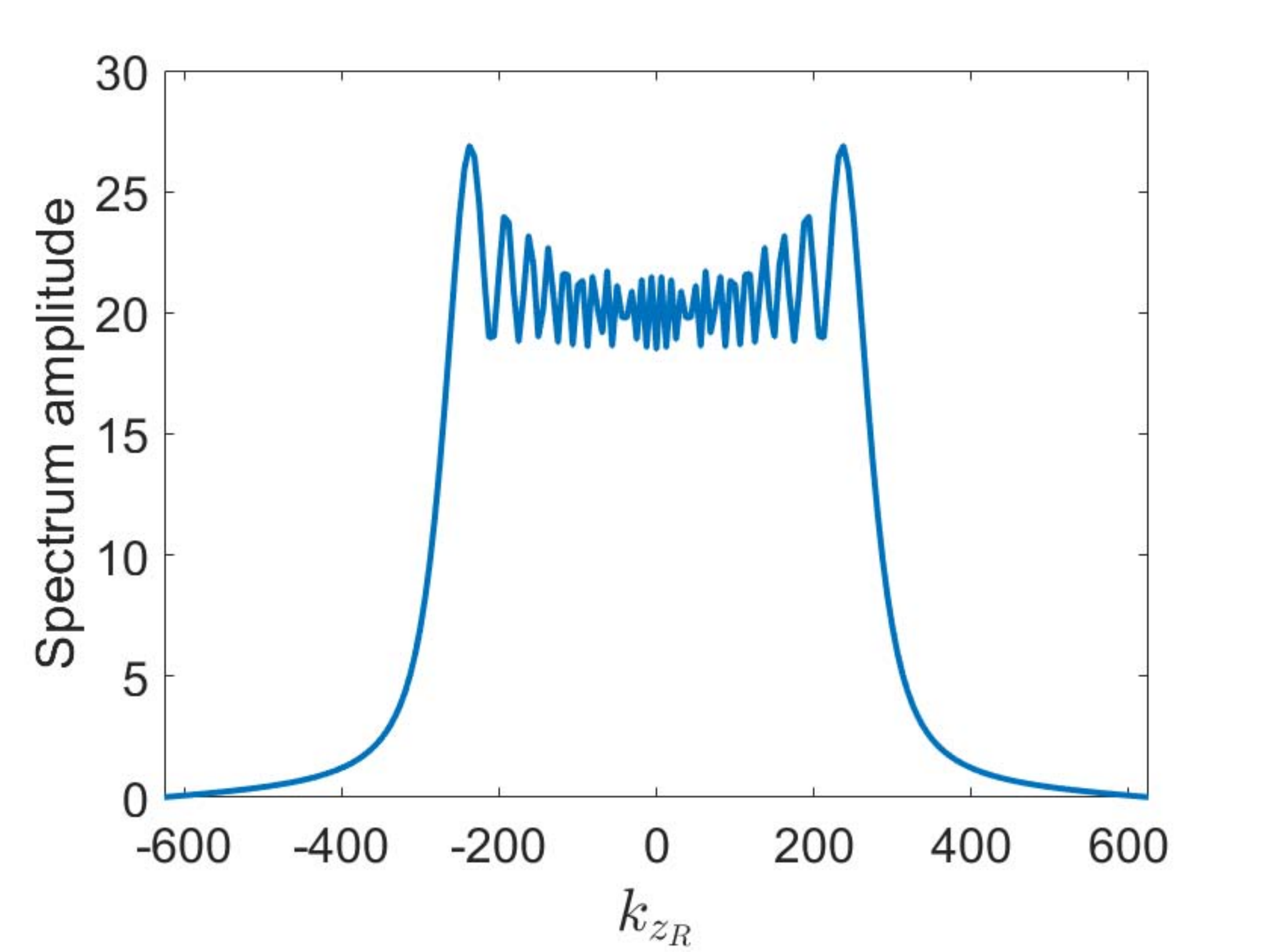}}
	\hfill
	\subfloat[]{\label{b}
		\includegraphics[width=1.69in]{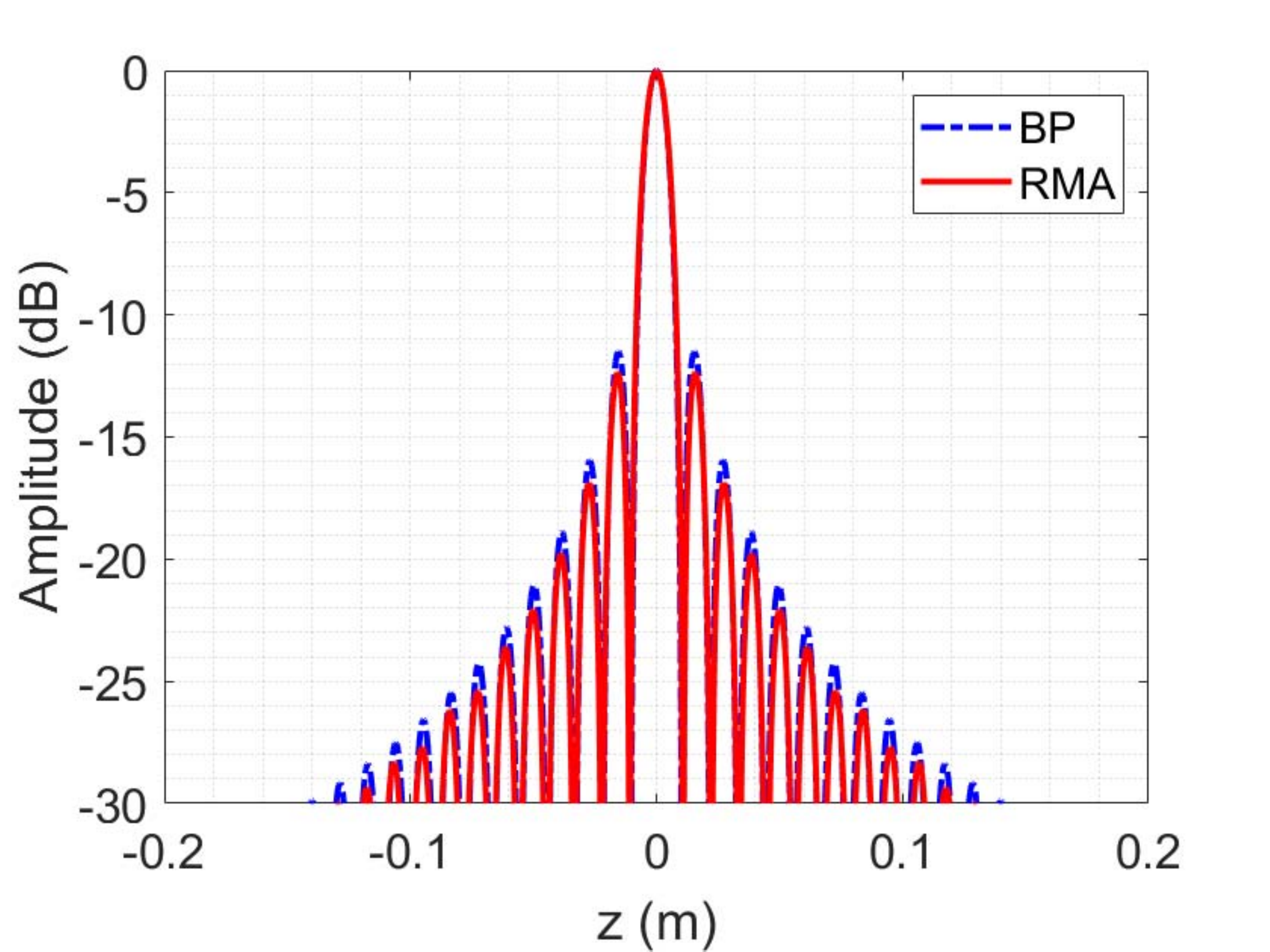}}
	\\
	
	\caption{(a) Spectrum and (b) near-field one-way beam pattern of the receive array.}
	\label{spect_psf_fullsamp}
\end{figure}

\begin{figure}[!t]
	\centering
	\subfloat[]{\label{a}
		\includegraphics[width=1.69in]{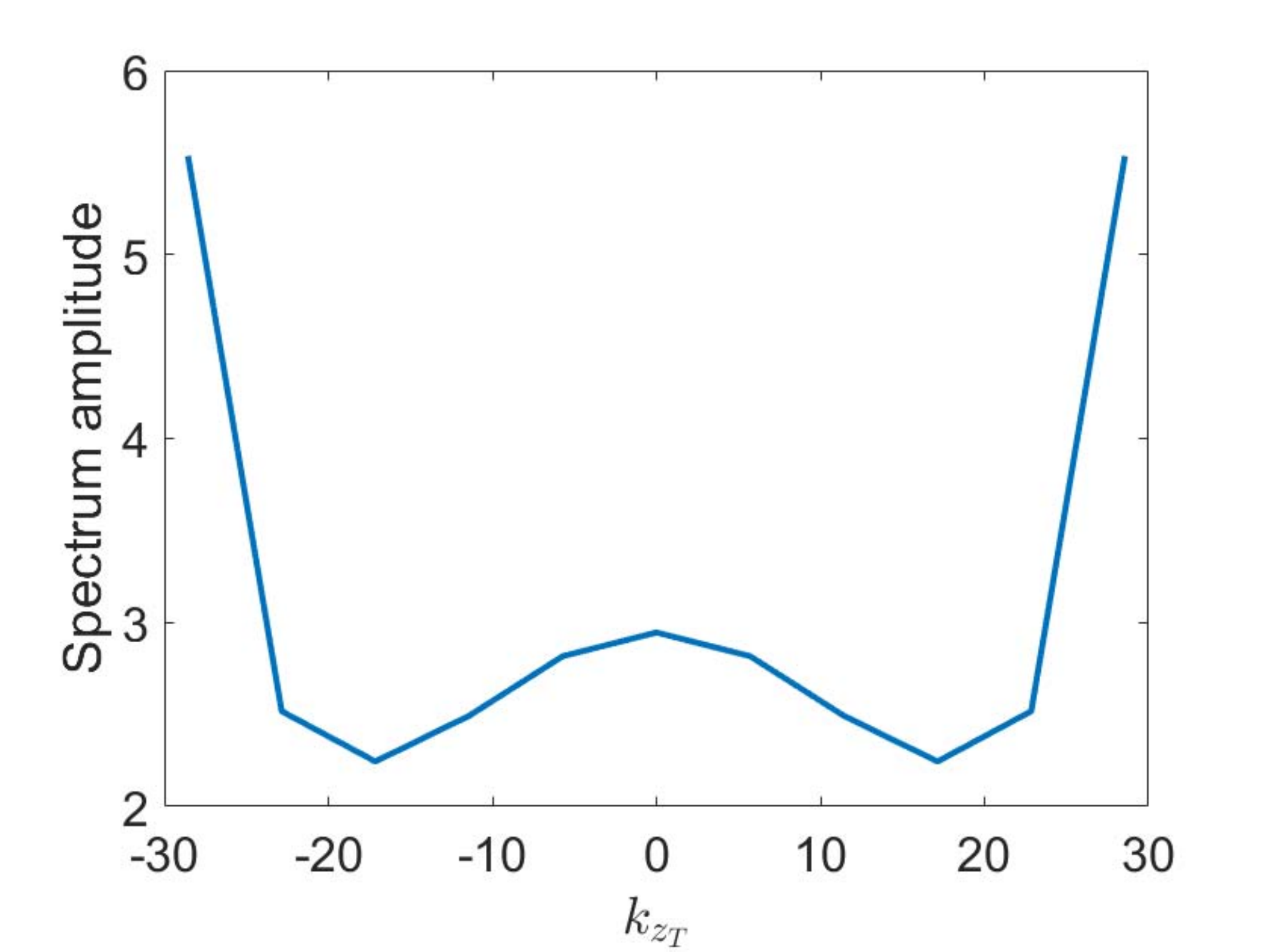}}
	\hfill
	\subfloat[]{\label{b}
		\includegraphics[width=1.69in]{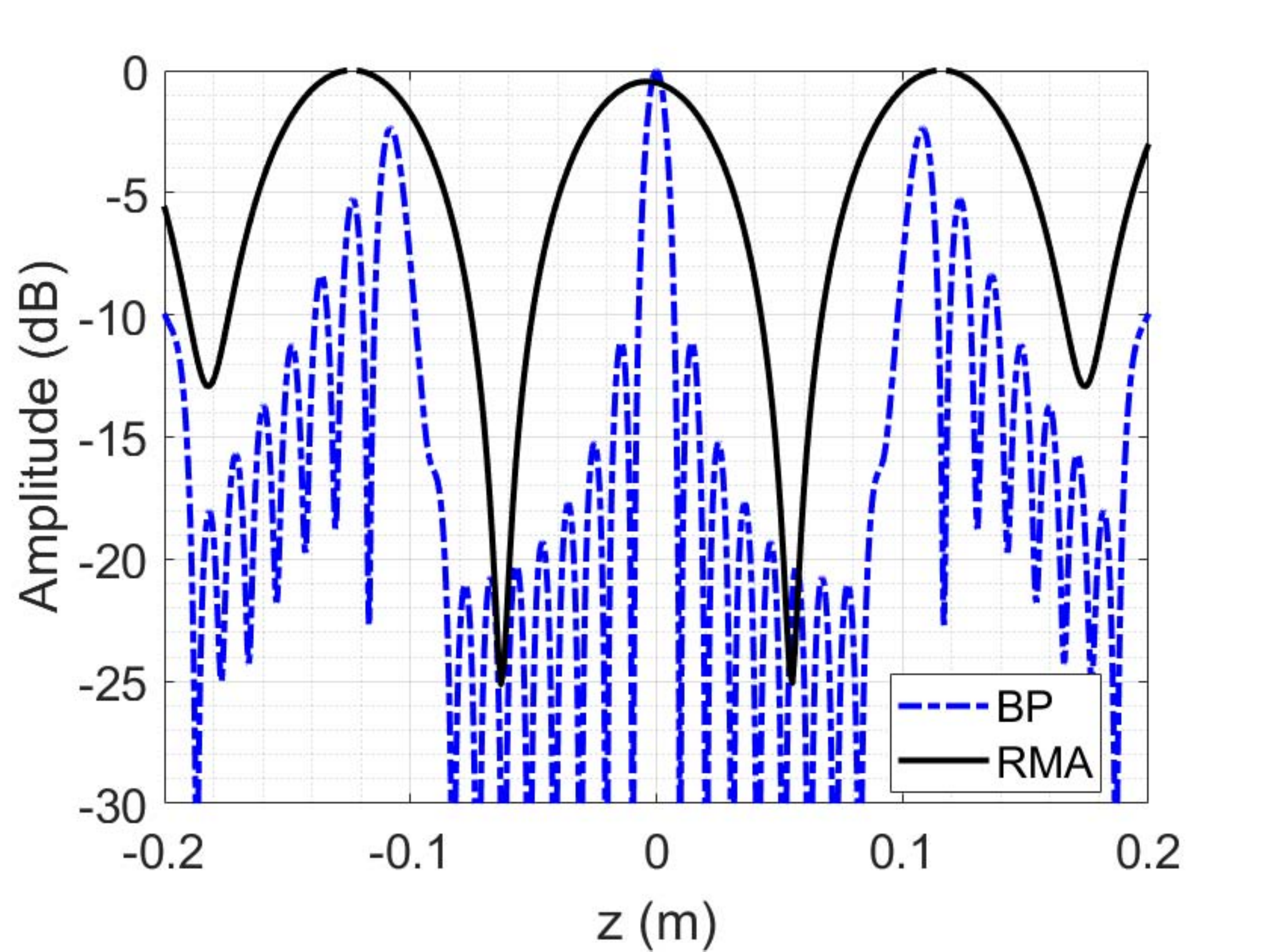}}
	\\
	
	\caption{(a) Spectrum and (b) near-field one-way beam pattern of the transmit array without zero padding.}
	\label{spect_psf_unsamp_no0}
\end{figure}

\begin{figure}[!t]
	\centering
	\subfloat[]{\label{a}
		\includegraphics[width=1.69in]{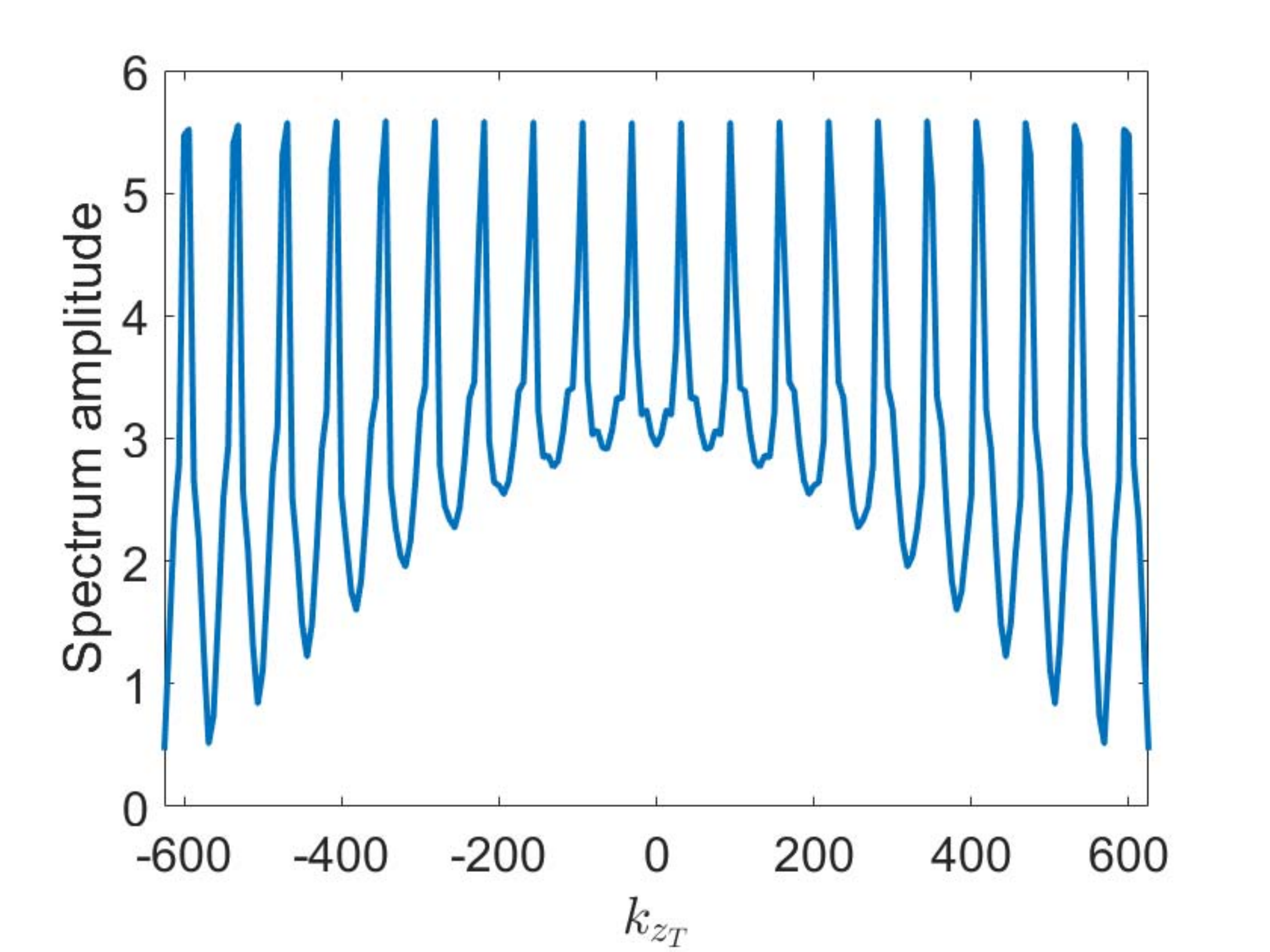}}
	\hfill
	\subfloat[]{\label{b}
		\includegraphics[width=1.69in]{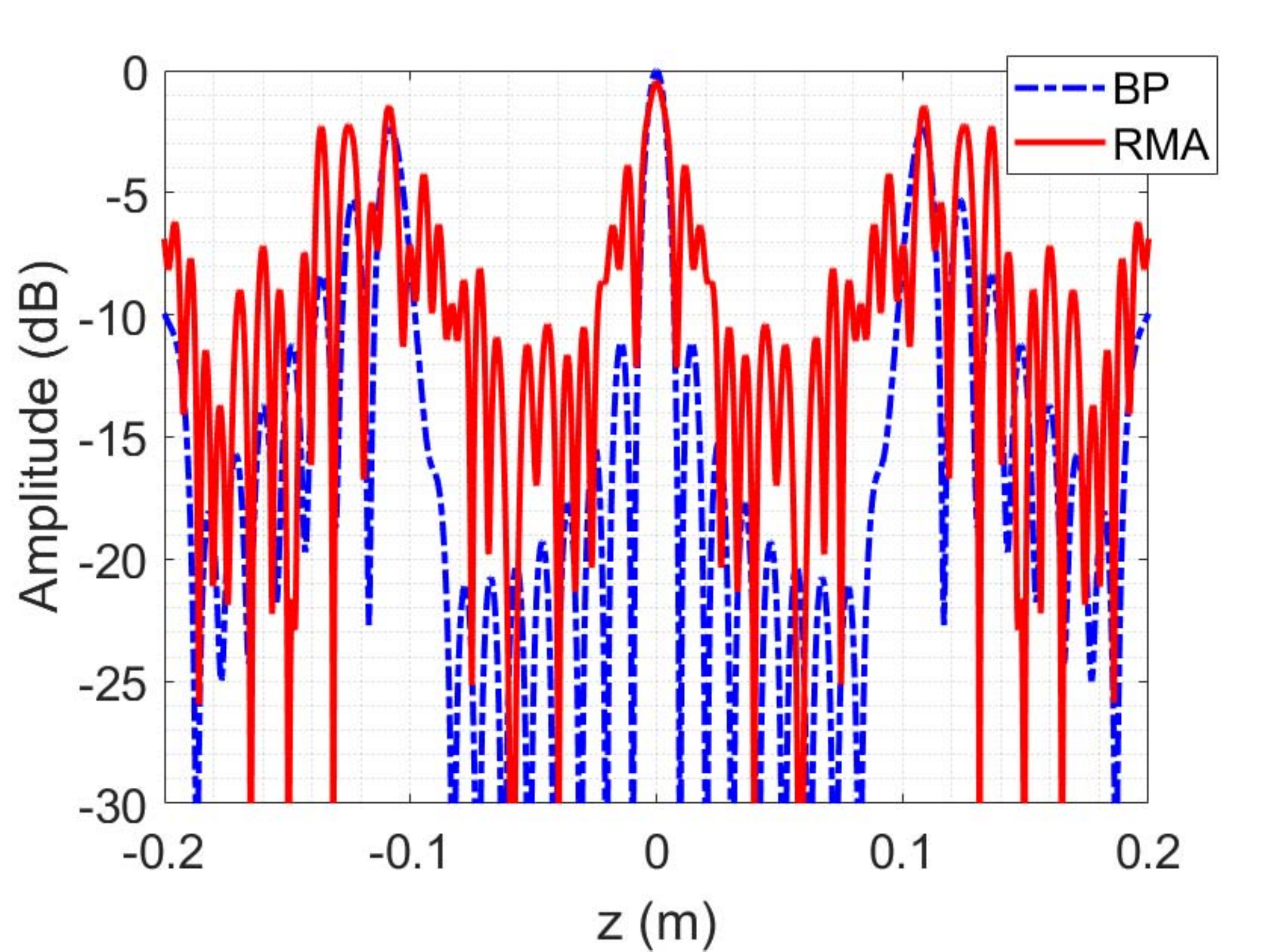}}
	\\
	
	\caption{(a) Spectrum and (b) near-field one-way beam pattern of the transmit array with zero padding.}
	\label{spect_psf_unsamp_0}
\end{figure}

\begin{figure}[!t]
	\centering
	\subfloat[]{\label{a}
		\includegraphics[width=1.69in]{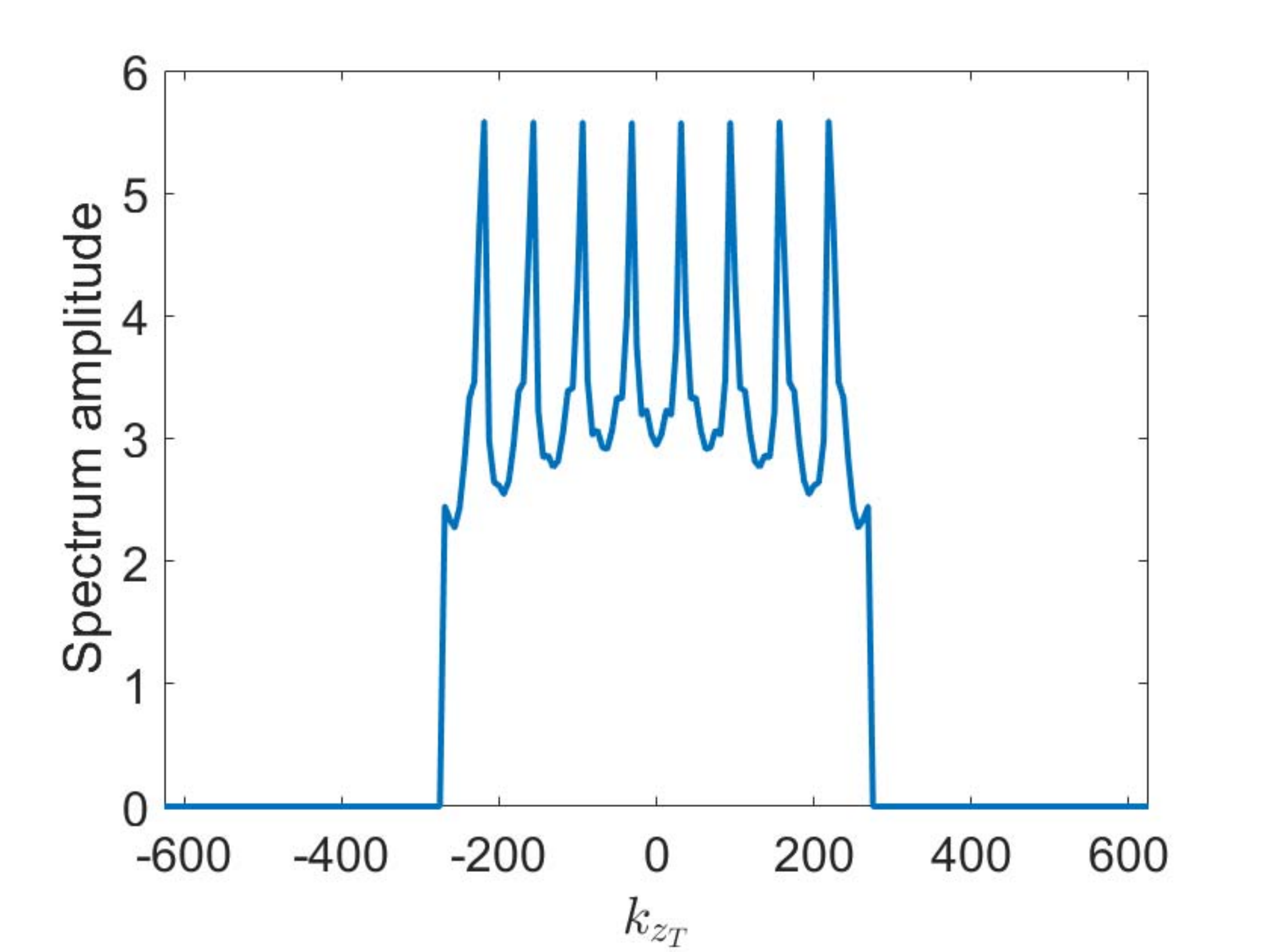}}
	\hfill
	\subfloat[]{\label{b}
		\includegraphics[width=1.69in]{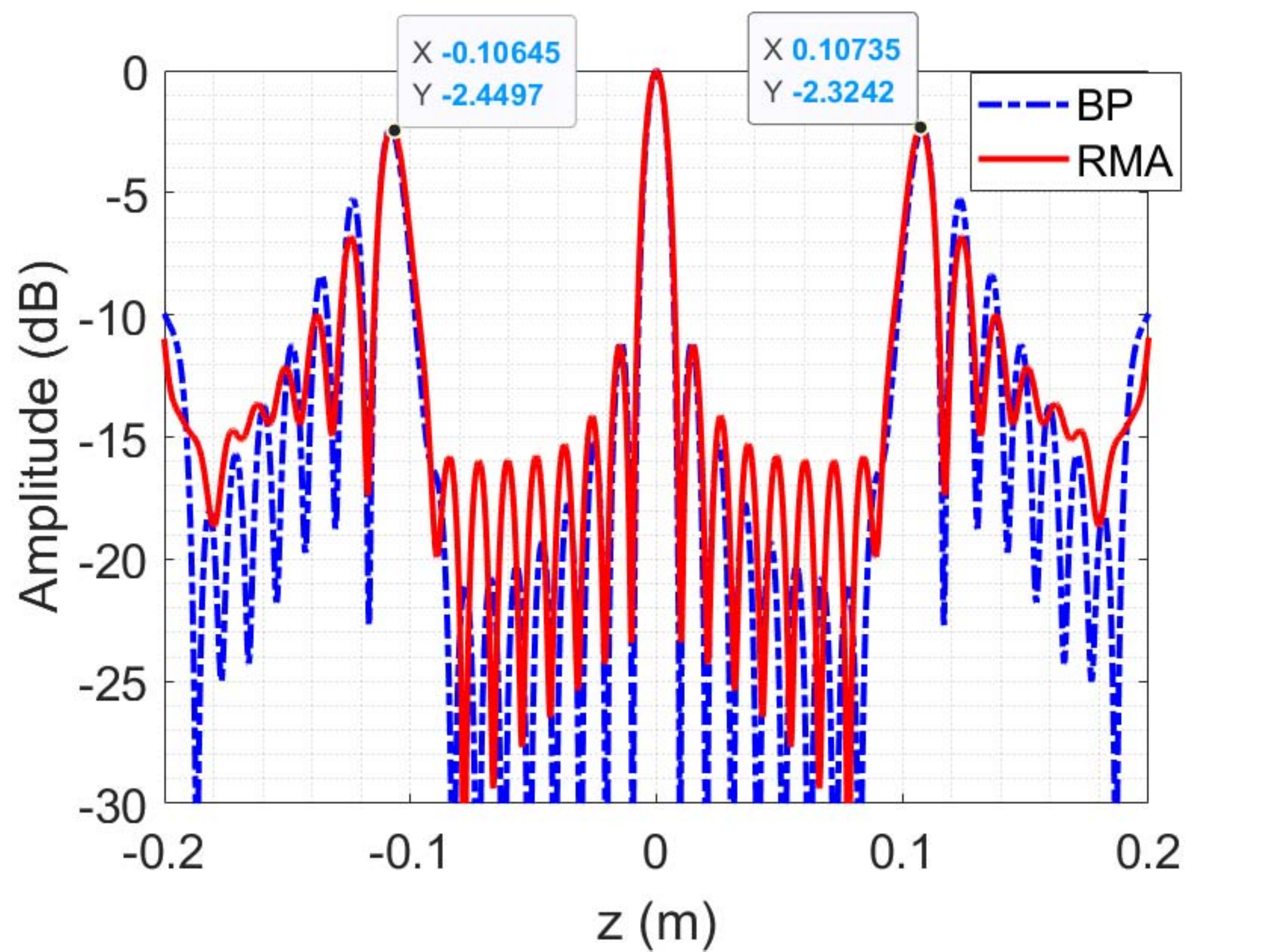}}
	\\
	
	\caption{(a) Filtered spectrum and (b) near-field one-way beam pattern of the transmit array with zero padding.}
	\label{spect_psf_unsamp_trunc}
\end{figure}

\begin{figure}[!t]
	\centering
	\includegraphics[width=2.5in]{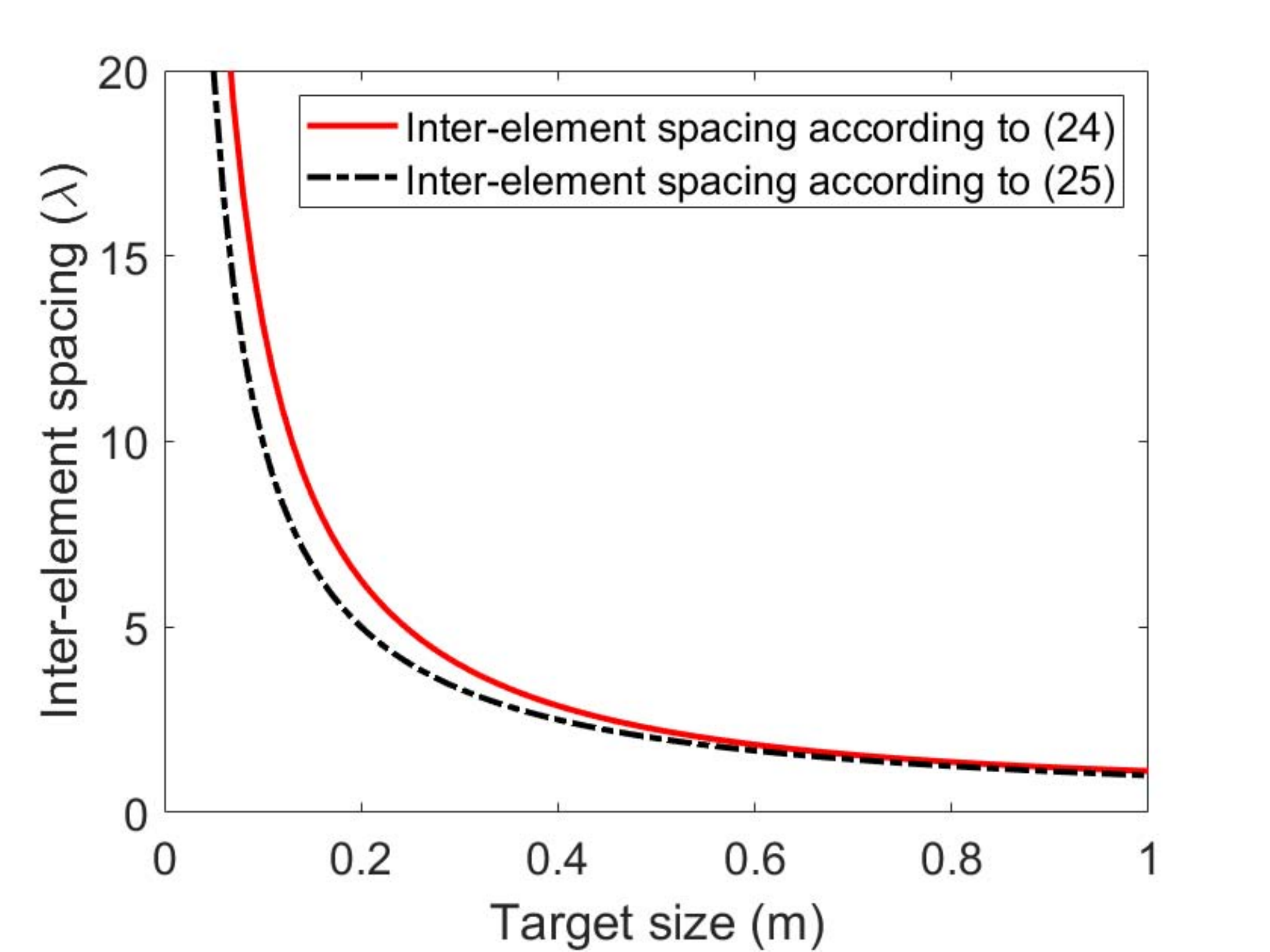}
	\caption{Comparison of the antenna intervals expressed by \eqref{un_intv} and \eqref{un_intv_appro}.}
	\label{comparison_antenna_interval_accu_appro}
\end{figure}

\begin{figure}[!t]
	\centering
	\subfloat[]{\label{a}
		\includegraphics[width=1.69in]{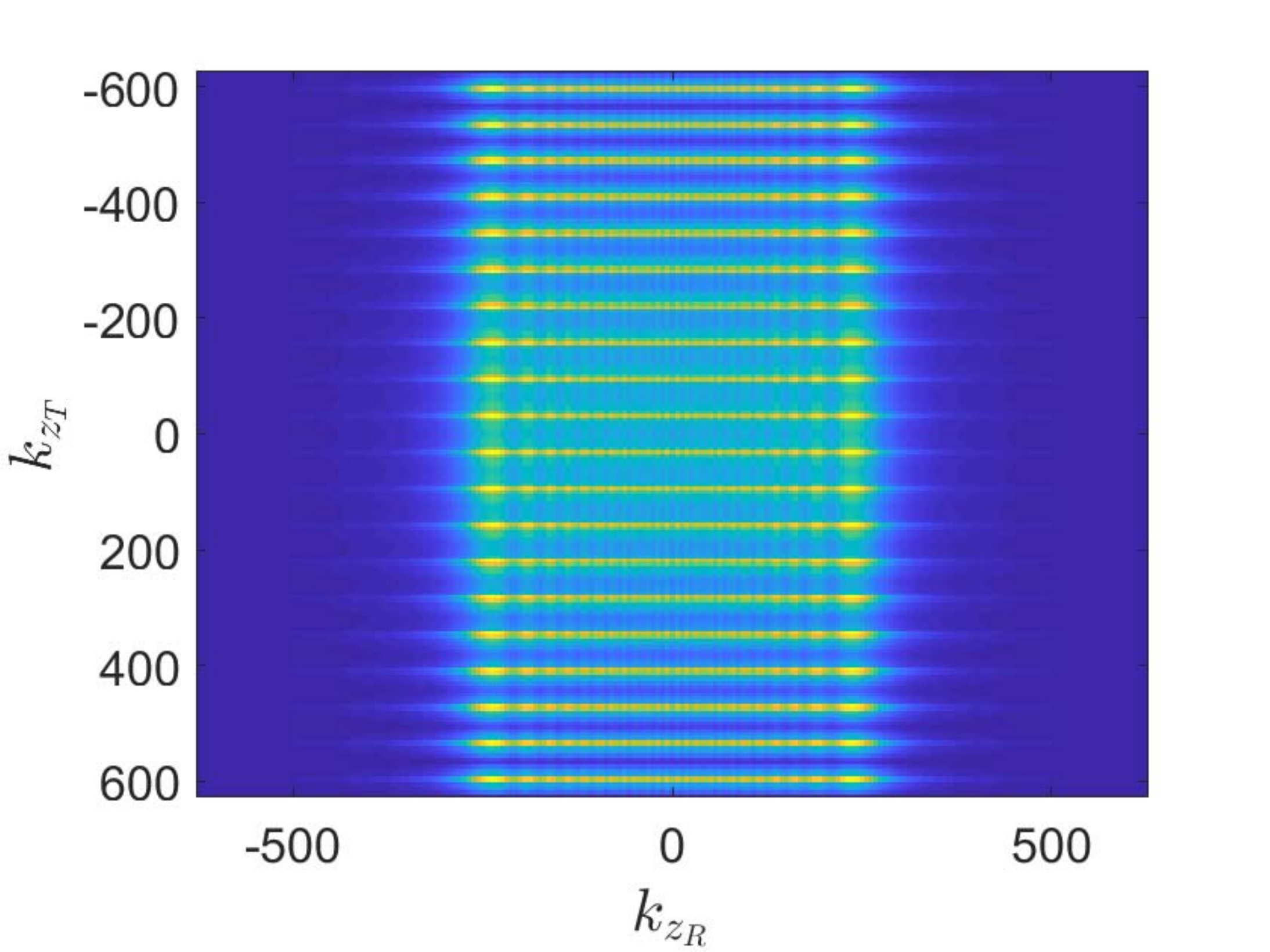}}
	\hfill
	\subfloat[]{\label{b}
		\includegraphics[width=1.69in]{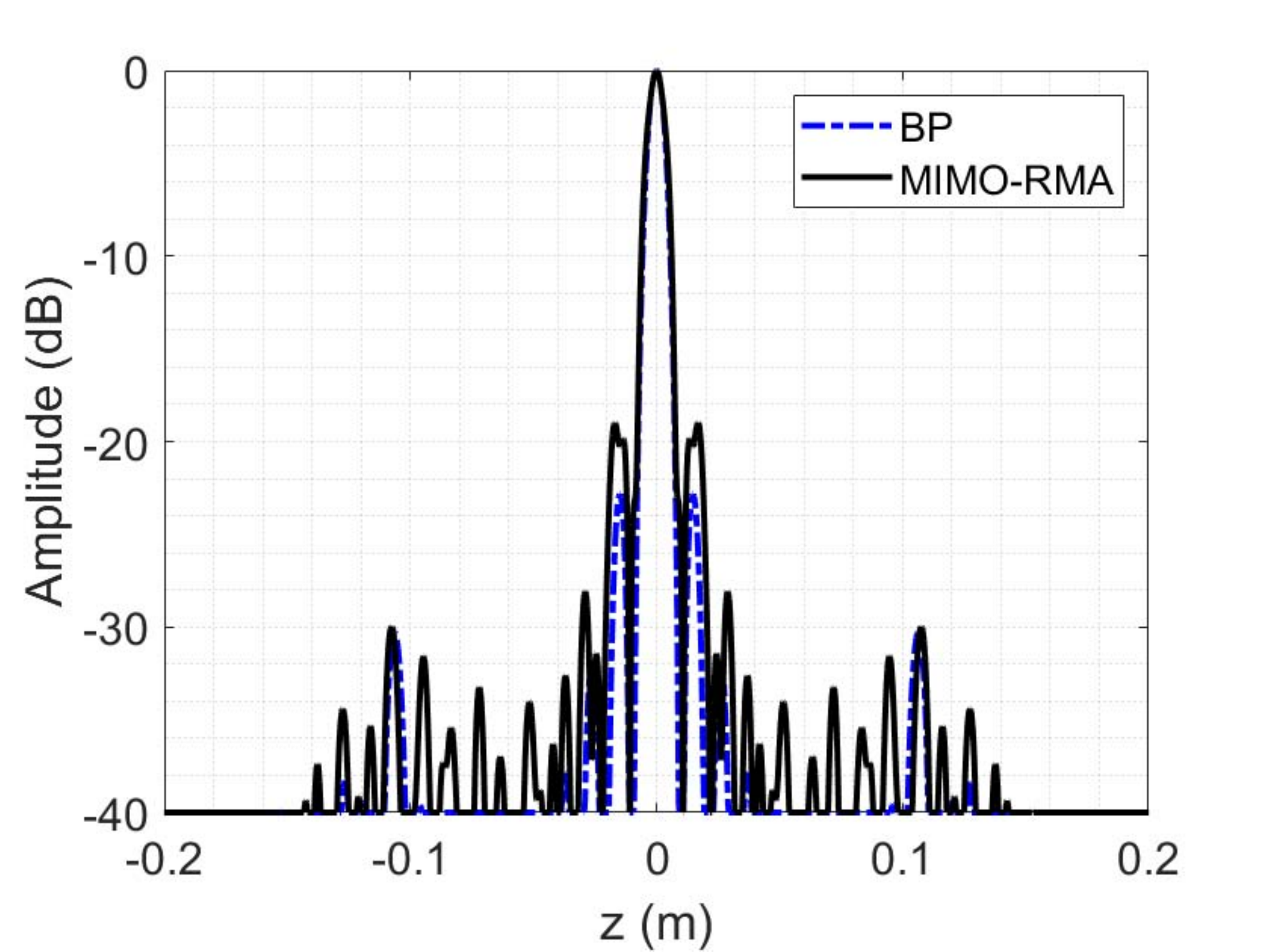}}
	\\
	
	\caption{(a) Unfiltered spectrum and (b) near-field two-way beam pattern of the MIMO array with zero padding.}
	\label{mimo_spect_psf_unsamp_untrunc}
\end{figure}

\begin{figure}[!t]
	\centering
	\subfloat[]{\label{a}
		\includegraphics[width=1.69in]{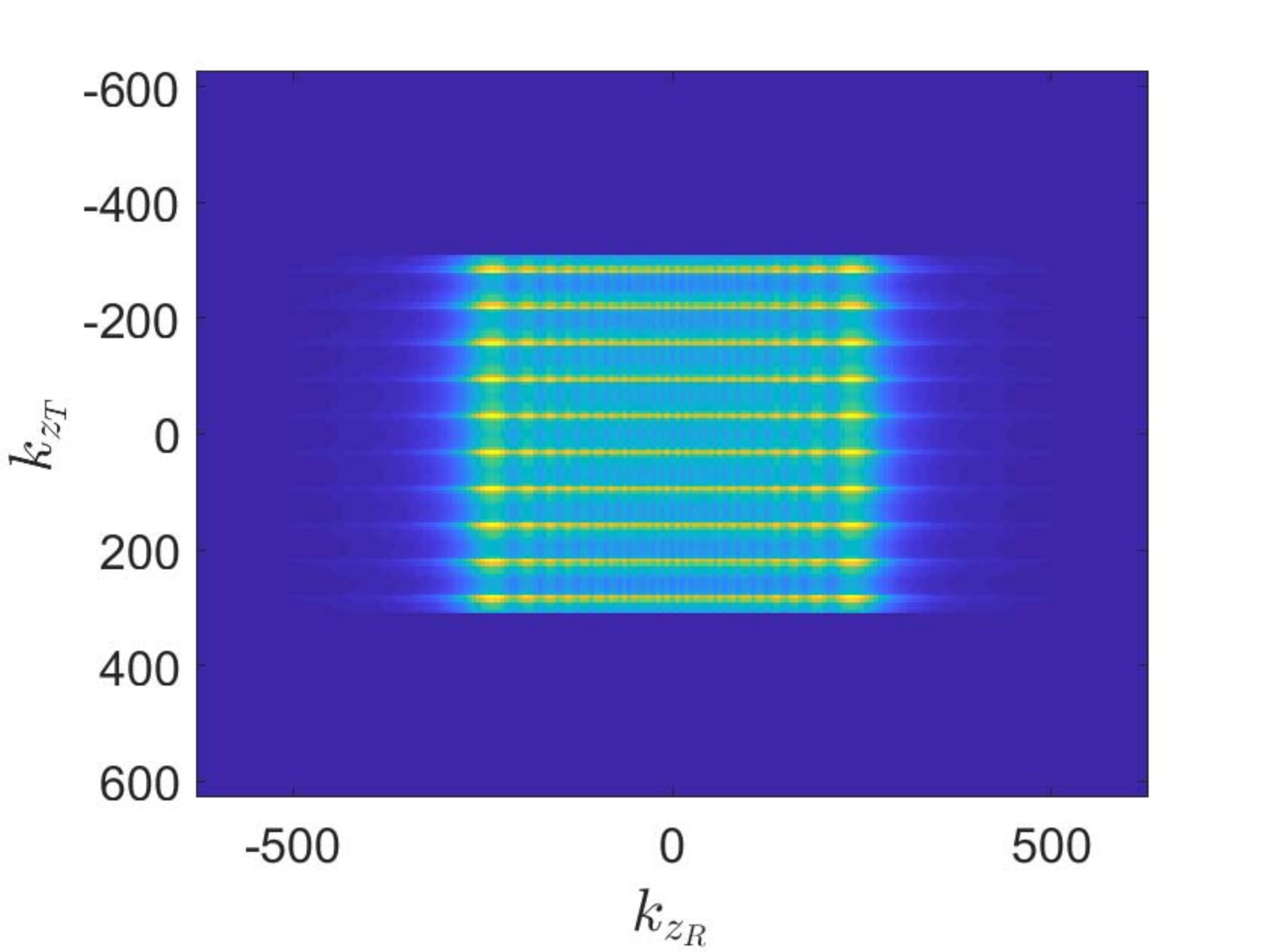}}
	\hfill
	\subfloat[]{\label{b}
		\includegraphics[width=1.69in]{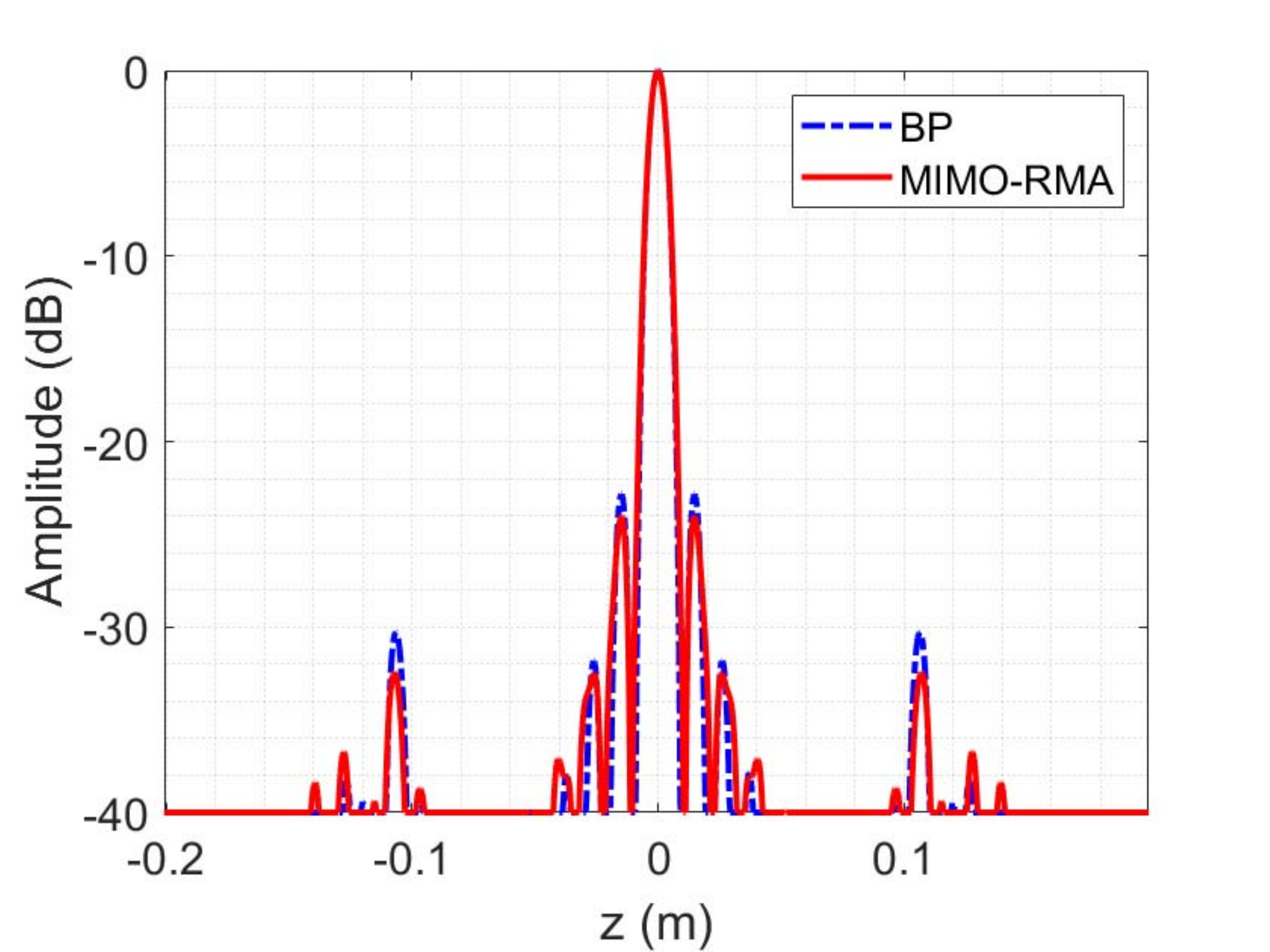}}
	\\
	\caption{(a) Filtered spectrum and (b) near-field two-way beam pattern of the MIMO array with zero padding.}
\label{mimo_spect_psf_unsamp_trunc}
\end{figure}

\begin{figure}[!t]
	\centering
	\includegraphics[width=2.5in]{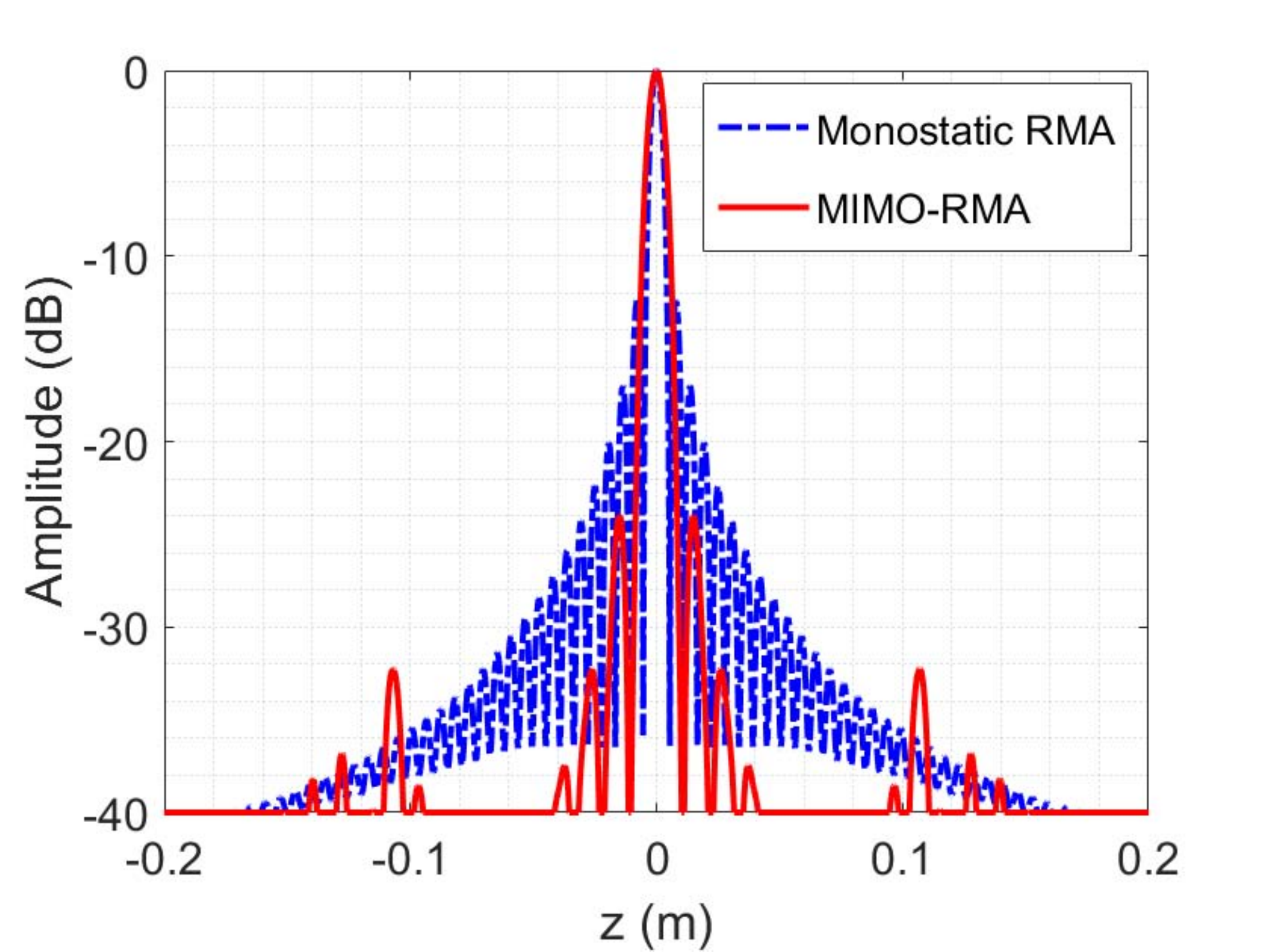}
	\caption{Comparison of the near-field beam patterns of a MIMO array with that of a monostatic array with a same aperture size.}
	\label{mimo_mono}
\end{figure}

The near-field beam patterns (evaluated at $R_0=1$m) are obtained by frequency-domain processing, referred to as the range migration algorithm (RMA)\cite{rma}. The corresponding results using BP are treated as benchmarks. 
The data in the spatial frequency domain  and the near-field beam pattern of the receive array (referred to as the fully sampled array) are shown in Figs. \ref{spect_psf_fullsamp}\subref{a} and \ref{spect_psf_fullsamp}\subref{b}, respectively. 
The related results of the transmit array (referred to as the undersampled array) without zero padding of the data are illustrated in Fig. \ref{spect_psf_unsamp_no0}. 
Note that only a small part of the original  spectrum  is obtained, which results in a beam pattern with a much wider mainlobe (different from a \textit{sinc} function) than the BP result.

The results, with $P-1$ zeros  placed between successive antenna elements of the transmit array to reach an equivalent inter-element spacing with the receive array, are shown in Fig. 
\ref{spect_psf_unsamp_0}. Clearly, the spectrum shows highly overlapping replicas. 
However, similar to Fig. \ref{spectrum_analysis}\subref{c}, there is a complete spectrum present in the result.  
The corresponding mainlobe of beam pattern is closer to that of BP  than  the result in Fig. \ref{spect_psf_unsamp_no0}\subref{b}.
But, the sidelobes are much higher than those of a standard \textit{sinc} function. 


To approach the BP result, we multiply the data in the spatial frequency domain   
by a rectangular window (in reality, it can be realized by using a weighted windowing function to reduce sidelobes), whose extent  is chosen  as the  extent of the true data spectrum that  can be calculated according to the relation between the wave vectors $\vec{k}$ and $\vec{k}_z$, in other words, based on 
the geometry between the array and the considered distance.
The  results are illustrated in Fig. \ref{spect_psf_unsamp_trunc}, where we can clearly identify the mainlobe at the center of the image zone, with slightly higher sidelobes than those of BP. The grating lobes are also very close to the BP results.

This equivalence can be deduced as follows.
Since  the interval between the successive spectra is $2\pi/\Delta z$ based on Fig. \ref{spectrum_analysis}\subref{c}. Then, from Fig. \ref{targets_spectra}, we obtain
\begin{align} \label{un_intv0}
	\frac{2\pi}{\Delta z}&=k_0(\sin\Theta_0 - \sin\Theta_1) \\ \nonumber
	&=k_0 \Bigg[\frac{L/2}{\sqrt{R_0^2+L^2/4}}-\frac{L/2-D}{\sqrt{R_0^2+(L/2-D)^2}}\Bigg],
\end{align}
where $k_0$ denotes the middle wavenumber of the working EM waves. Here, we consider the distance between the true target and the virtual one in Fig. \ref{targets_spectra}\subref{a} as  $D$.  Accordingly, the  spacing between successive antennas of the undersampled array is given by,
\begin{equation} \label{un_intv}
	\Delta z=\lambda_0\frac{1}{\frac{L/2}{\sqrt{R_0^2+L^2/4}}-\frac{L/2-D}{\sqrt{R_0^2+(L/2-D)^2}}}. 
\end{equation}
where $\lambda_0=2\pi/k_0$ denotes the middle wavelength. 
If we use $\tan\Theta$ to relace $\sin\Theta$ to simplify \eqref{un_intv0}, we obtain an approximate expression of \eqref{un_intv},
\begin{equation} \label{un_intv_appro}
	\Delta z=\frac{\lambda_0 R_0}{D}, 
\end{equation}
which is clearly consistent with that of  BP.
 
Fig. \ref{comparison_antenna_interval_accu_appro} shows a comparison between \eqref{un_intv_appro} and \eqref{un_intv} (the value is normalized by $\lambda_0$), assuming $L=1$m, $R_0=1$m, and $D$ varying from 0 to 1m.  It is evident that the two curves match pretty well.  On the other hand, for a given $\Delta z$, $\lambda_0$ and $R_0$, we can calculate the distance $D$ between the mainlobe and the grating lobe. For example, when we choose the same parameters of Figs. \ref{spect_psf_fullsamp} to \ref{spect_psf_unsamp_trunc}, (i.e., $\Delta z=0.1$m, $\lambda_0=1$cm, and $R_0=1$m), the value of $D$ calculated by \eqref{un_intv_appro} is equal to 0.1m, which is very close to the one marked in Fig. \ref{spect_psf_unsamp_trunc}\subref{b}.

Finally, we show in Figs. \ref{mimo_spect_psf_unsamp_untrunc} and \ref{mimo_spect_psf_unsamp_trunc} the results of the MIMO array  by combining the transmit and the receive arrays. 
Due to multiplication of the transmit and the receive beam patterns, the amplitudes of the grating lobes in 
Figs. \ref{spect_psf_unsamp_0} and \ref{spect_psf_unsamp_trunc} are reduced to  lower than -30dB. 

Fig. \ref{mimo_mono} shows the comparison between the MIMO configuration  and that of 
a monostatic array with a same aperture size.
Note that the resolution of the proposed technique for MIMO array is only mildly coarser  than that of a monostatic array (see Table \ref{tab1}). 
This is caused by the convolution between the transmit and receive spectra. In essence, the beam pattern of the MIMO array has a wider mainlobe, but  lower sidelobes than those of the monostatic array.  






The resolutions and peak sidelobe ratios (PSLRs) of the different scenarios are listed in Table \ref{tab1}. 
Note that these results are consistent with the spectrum aliasing analysis. 
With regards to the MIMO case, the resolution and PSLR obtained by the proposed technique are comparable to those of the BP algorithm.

\begin{table}
	\centering
	\caption{Comparison of Resolutions and PSLRs of Beam Patterns}
	\label{tab1}
	\setlength{\tabcolsep}{3pt}
	\begin{threeparttable}
		\begin{tabular}{p{170pt}  p{35pt}  p{22pt}}
			\hline\hline
			Scenarios and methods& Resolutions (mm)& PSLRs (dB) \\[0.5ex]
			\hline
			fully sampled array by BP& 9.69& -11.51 \\[0.5ex]
			fully sampled array by RMA& 9.74& -12.36 \\[0.5ex]
			undersampled array by RMA without zero filling& 59.8&  $\sim$ \\[0.5ex]
			undersampled array by RMA with zero filling but without spectrum filtering&  10.70&  -3.44\\[0.5ex]
			undersampled array by RMA with zero filling and spectrum filtering& 9.25& -11.24 \\[0.5ex]
			MIMO array by BP& 6.56& -22.82 \\[0.5ex]			
			MIMO array by RMA without spectrum filtering& 7.21& -18.98 \\[0.5ex]
			MIMO array by RMA with  spectrum filtering& 6.66& -24.01 \\[0.5ex]
			Monostatic array by RMA& 4.87& -12.40 \\[0.5ex]			
			\hline
		\end{tabular}
	\end{threeparttable}
\end{table}

\subsection{Sampling Criteria}

\begin{figure}[!t]
	\centering
	\includegraphics[width=1.5in]{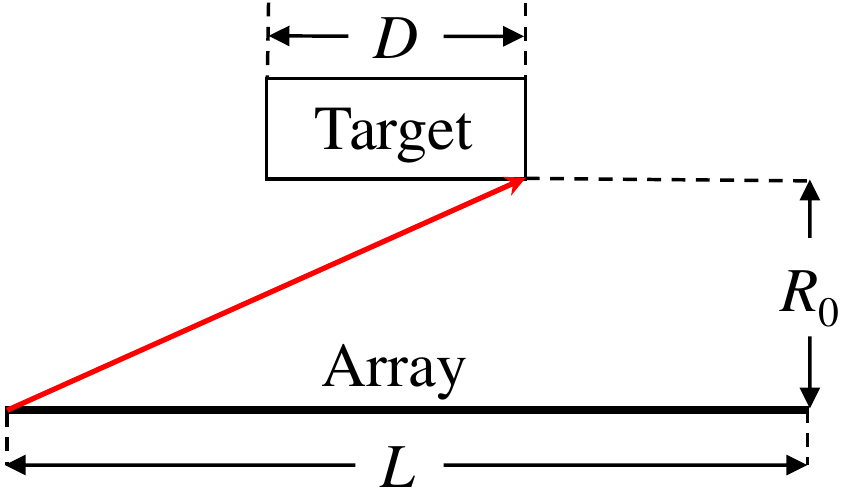}
	\caption{Geometrical information for computing the Nyquist sampling interval of the transmit or receive array.}
	\label{antenna_interval_fullsamp}
\end{figure}

From the above analysis, it is apparent  that the MIMO array combining the fully sampled array and the undersampled array can achieve good imaging result. Next, we present the corresponding sampling requirements for both arrays. 
According to $e^{-\text{j}k_z z}$, the inter-element spacing $\Delta z$ of the fully sampled array should satisfy $2\pi/\Delta z \geq k_{z_\text{max}}-k_{z_\text{min}}$ to ensure no spectrum aliasing would occur. Based on the linear array shown in Fig. \ref{antenna_interval_fullsamp}, we obtain 
\begin{equation} \label{nyquist_intv}
	\Delta z\leq \lambda_0 \frac{\sqrt{R_0^2+{(L+D)^2}/{4}}}{L+D}
\end{equation}
which is consistent with the requirement in \cite{zhuge2,1d_mimo_cylindrical}.
Here,  $L$ denotes the array length, and $D$ represents the target dimension along the $z$ direction.

In regards of the  antenna spacing associated with the  undersampled array, we note that the final imaging result of MIMO is equivalent to the multiplication of the result of the transmit array by that of the receive array. Thus, the inter-element spacing of the undersampled array does not need to satisfy  \eqref{nyquist_intv}, as long as we process the data using the method mentioned in Sec. III. B. Theoretically, the MIMO array can achieve a similar resolution with the monostatic one, even with only two antennas for the undersampled subarray, which are located at the both ends of the fully sampled subarray.

The maximal angular interval $\Delta\theta$ of the antennas for the fully sampled array along the horizontal direction should meet $k_{\rho}\sin\Delta \theta\approx k\sin\Delta \theta \leq 2\pi/D$, that yields,
\begin{equation}
\Delta \theta \leq \frac{\lambda_0}{D},
\end{equation}
just as the same for the requirement in BP. Here, $D$ denotes the maximal dimension of the target along the horizontal direction. 
Similarly, the  undersampled array  can exceed this requirement.  





\subsection{Resolutions}

First, we consider the cross-range resolution along the horizontal direction, i.e., the resolution of the circular MIMO array. Due to the exponential function $\exp(-\mathrm{j}k_{x}x)$, the  cross-range resolution $\delta x$ is determined by the range of the spatial frequency $k_x$, that is, 
\begin{equation}
	\delta x = \frac{\pi}{k_{x_{\text{max}}}},
\end{equation}
where $k_{x_{\text{max}}}$ denotes the maximum value of $k_x$.

According to the relation $k_x=k_{x_T}+k_{x_R}$, and the array configuration in Fig. \ref{cylindrical_mimo}, we note that $k_{x_{\text{max}}}=k_{x_{T_{\text{max}}}}+k_{x_{R_{\text{max}}}}$ with $k_{x_{T_{\text{max}}}}=k_{x_{R_{\text{max}}}}\approx k_c\sin(\Theta_h /2)$ with the help of \eqref{krhot}, \eqref{krhor}, and \eqref{kxtr}, where $k_c$ denotes the center wavenumber of the working EM waves, and $\Theta_h$ represents the maximal angle subtended by the array aperture,
assuming the beamwidth of each antenna can fully illuminate the target in the horizontal direction. Thus, we have
\begin{equation}
	\delta x = \frac{\pi}{2k_c\sin \frac{\Theta_h}{2}} = \frac{\lambda_c}{4\sin\frac{\Theta_h}{2}}.
\end{equation}
This is the same result for  the monostatic imaging scenario. 

Correspondingly, the resolution along the vertical direction is given by,
\begin{equation}
	\delta z = \frac{\pi}{k_{z_{\text{max}}}}=\frac{\lambda_c}{4\sin\frac{\Theta_z}{2}},
\end{equation}
where $\Theta_z$ denotes the minimal value of the angle subtended by the vertical array length from the target and the beamwidth of the antenna element. Clearly, it is also the same as the monostatic one if we design the transmit and receive arrays with a same aperture size.


Finally, the down-range resolution is determined by, 
\begin{equation}
\delta y = \frac{c}{2B},
\end{equation}
where $B$ represents the bandwidth of the working EM waves.




\section{Results}

\begin{table}
	\centering
	\caption{Simulation Parameters}
	\label{tab3}
	\setlength{\tabcolsep}{3pt}
	\begin{threeparttable}
		\begin{tabular}{p{200pt}  p{30pt}}
			\hline\hline
			Parameters& Values \\[0.5ex]
			\hline
			Radius of the cylindrical aperture $(R_0)$&
			1.5 m\\[0.5ex]
			Start frequency& 
			31 GHz \\[0.5ex]
			Stop frequency&
			39 GHz \\[0.5ex]
			Number of frequency steps&
			15 \\[0.5ex]
			Inter-element spacing of transmit array along elevation&
			10 cm \\[0.5ex]

			Inter-element spacing of receive array along elevation&
			1 cm \\[0.5ex]
			Number of transmit antenna elements along elevation&
            5 \\[0.5ex]			
			Number of receive antenna elements along elevation&
			41 \\[0.5ex]
			Inter-element spacing of  transmit array along circumference&
			9.9 cm \\[0.5ex]
			Inter-element spacing of receive array along circumference&
			0.99 cm \\[0.5ex]
			Number of transmit antenna elements along circumference&
			5 \\[0.5ex]
			Number of receive antenna elements along circumference&
			41 \\[0.5ex]

			\hline
		\end{tabular}
		\label{tab2}
	\end{threeparttable}
\end{table}

\begin{figure}[!t]
	\centering
	\includegraphics[width=2.5in]{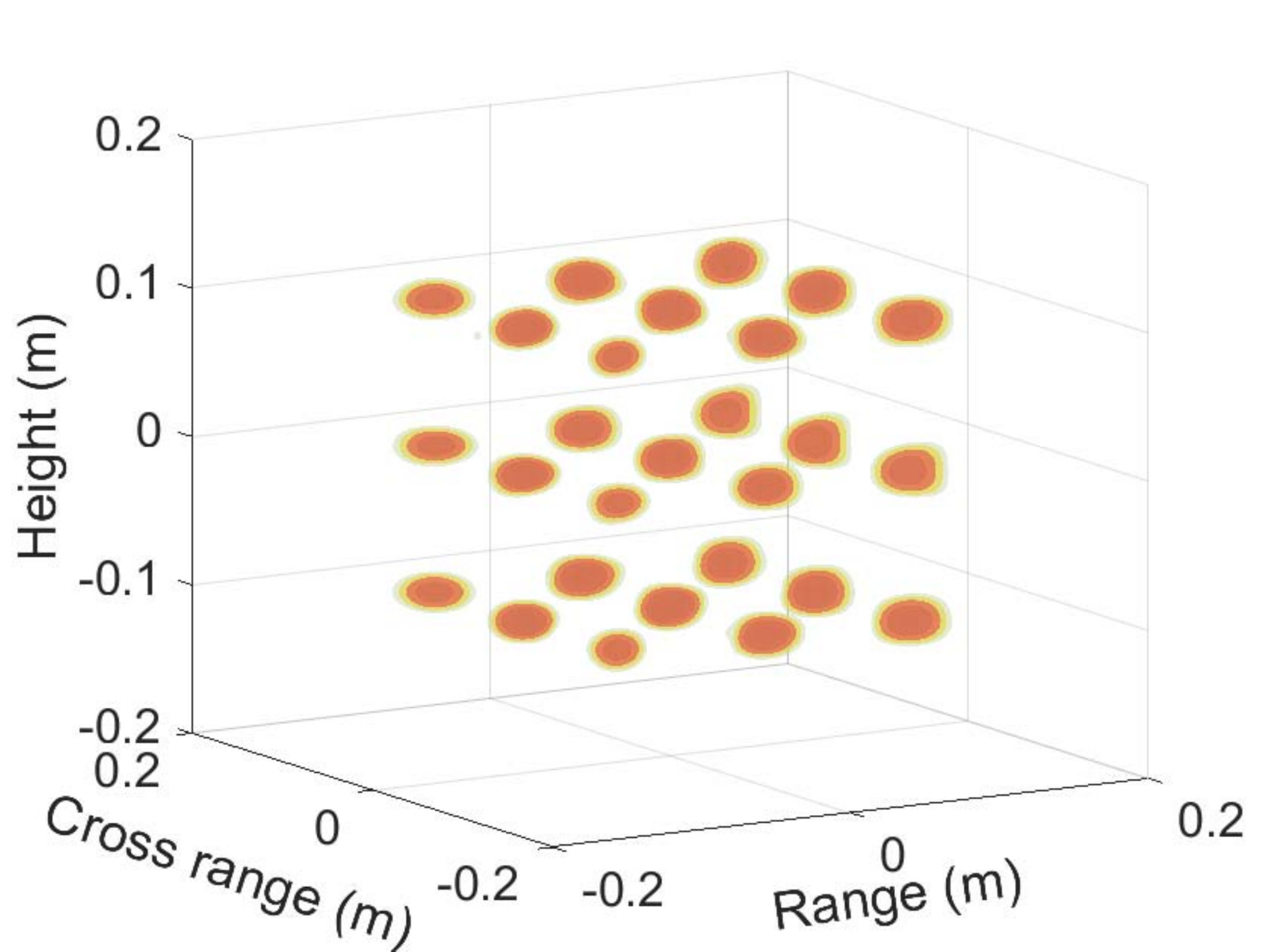}
	\caption{3-D Imaging result of the cylindrical RMA.}
	\label{3d_mimo}
\end{figure}

\begin{figure}[!t]
	\centering
	\subfloat[]{\label{a}
	\includegraphics[width=2.5in]{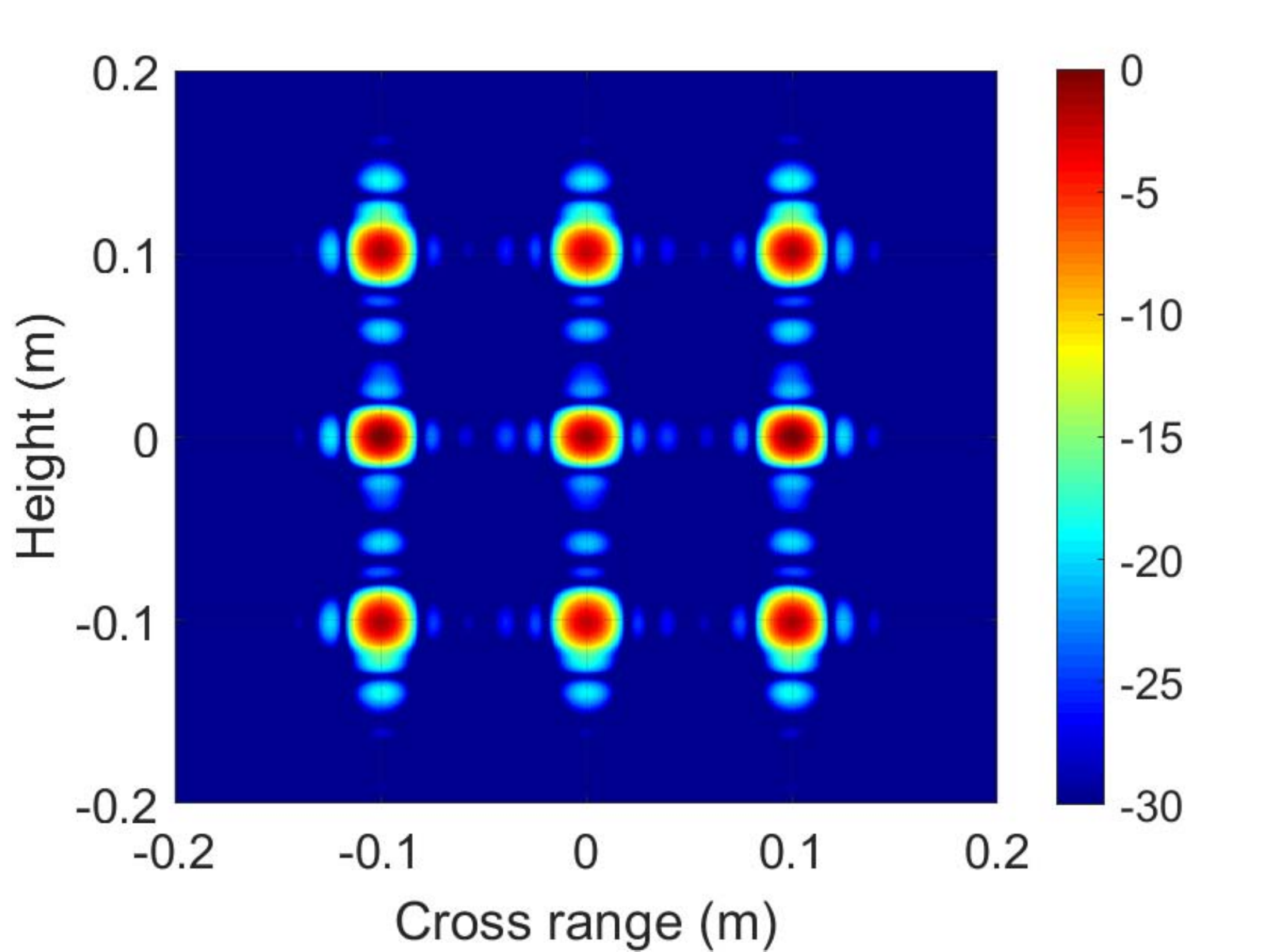}}
	\hfill
	\subfloat[]{\label{b}
	\includegraphics[width=2.5in]{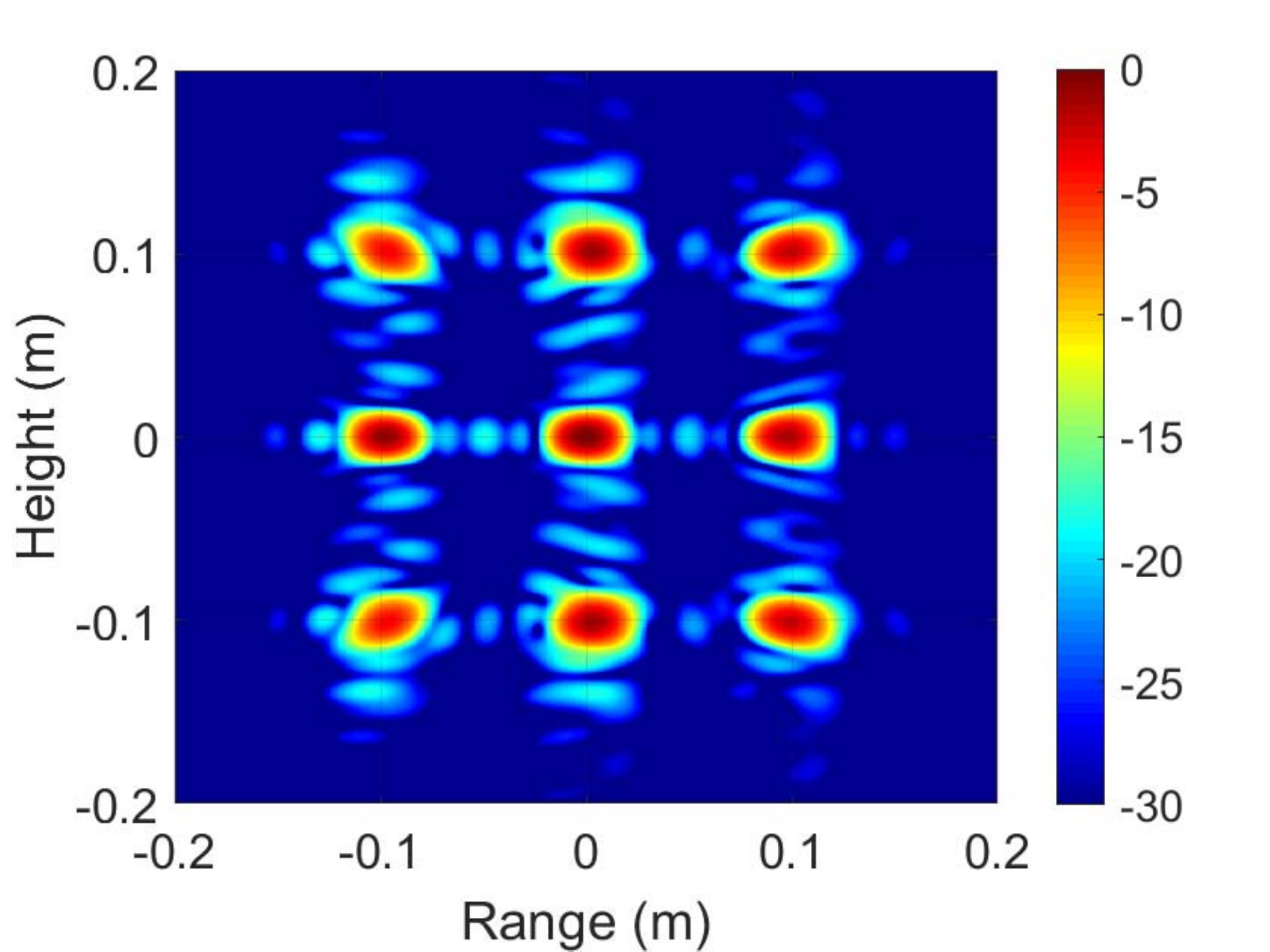}}
	\hfill
	\subfloat[]{\label{c}	
	\includegraphics[width=2.5in]{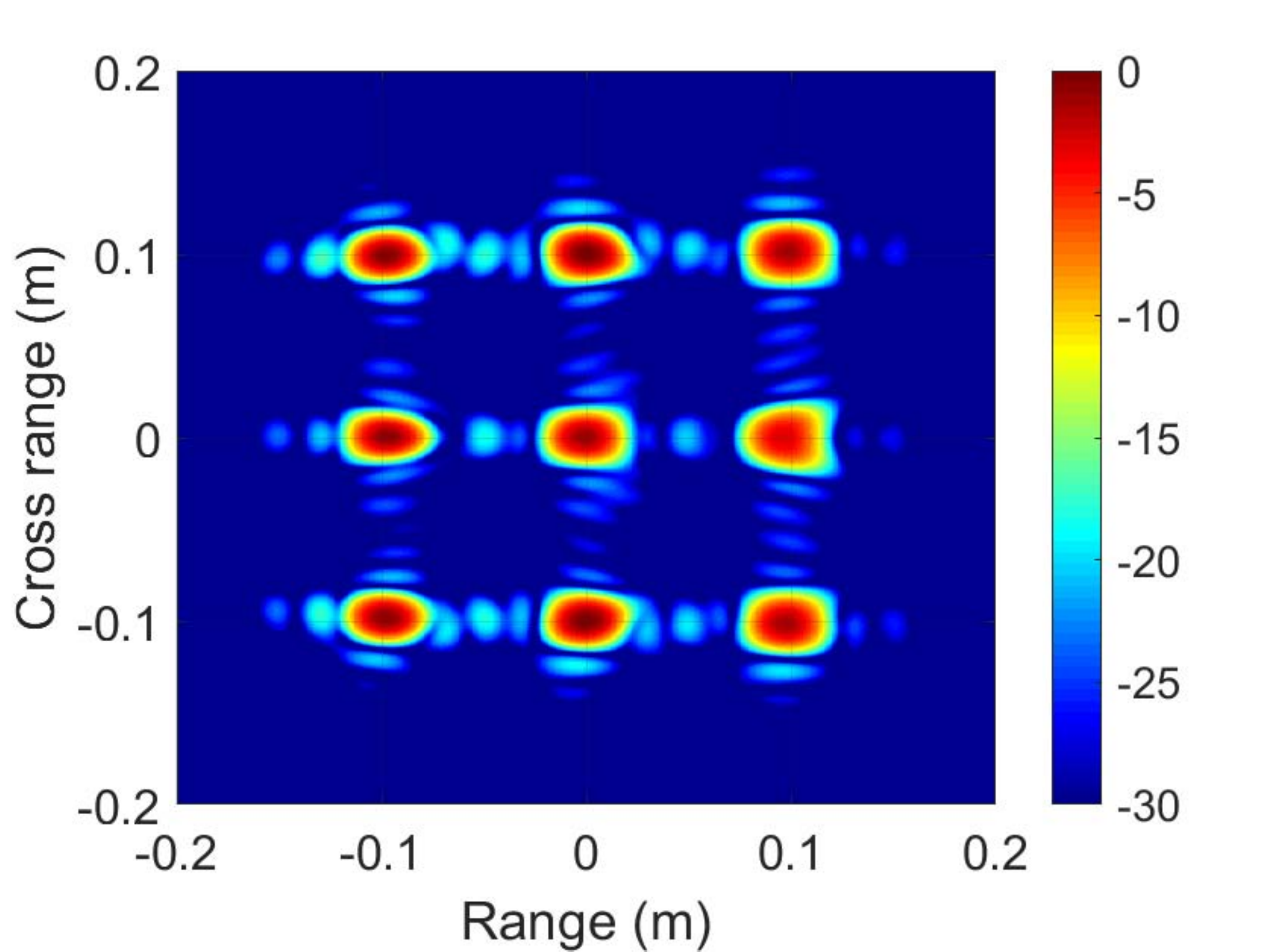}}	
	\\	
	\caption{2-D image slices corresponding to the (a) height-horizontal, (b) height-range, and (c) horizontal-range planes, respectively.}
	\label{2-D slices}
\end{figure}

\begin{figure}[!t]
	\centering
	\subfloat[]{\label{a}
		\includegraphics[width=2.5in]{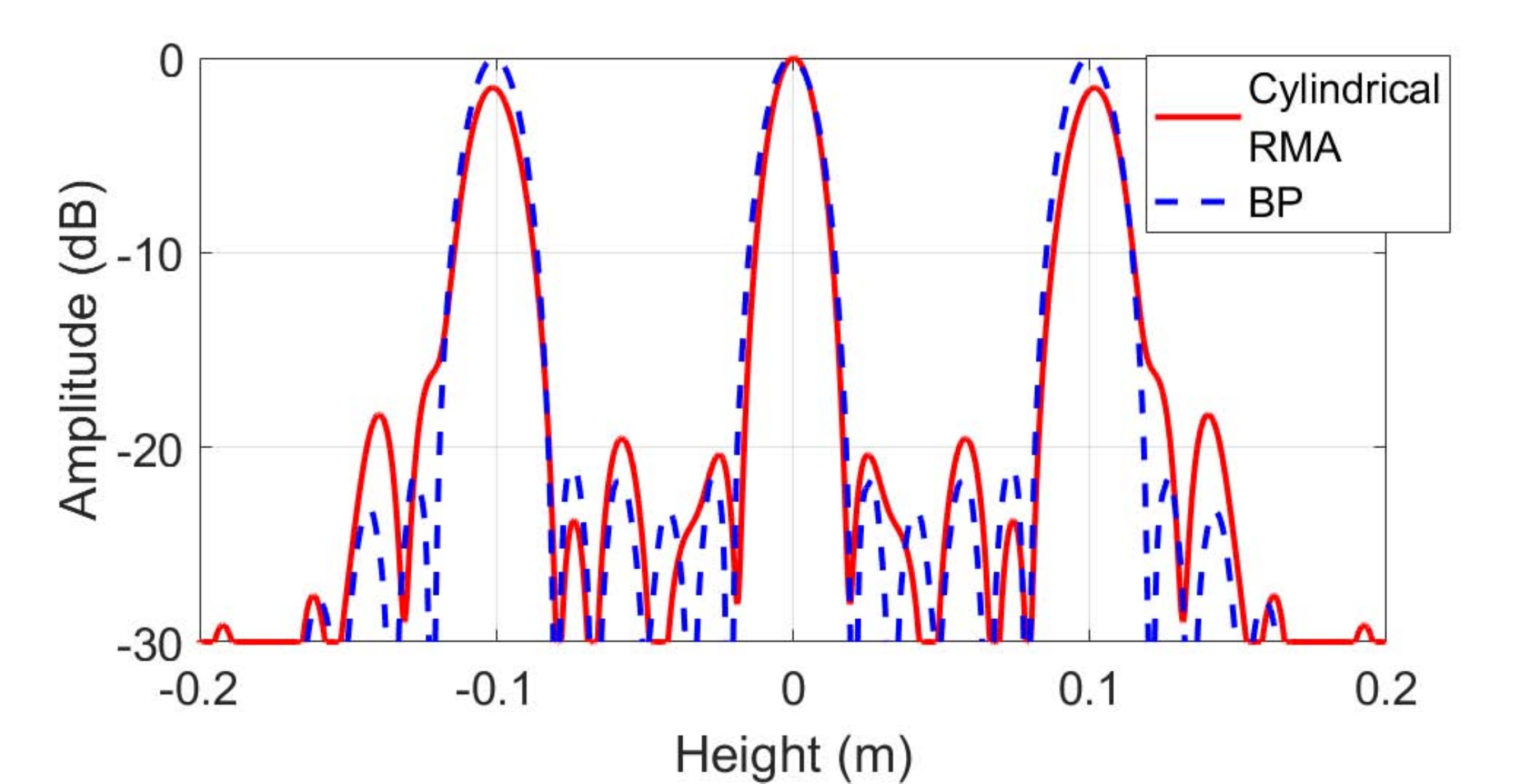}}
	\hfill
	\subfloat[]{\label{b}
		\includegraphics[width=2.5in]{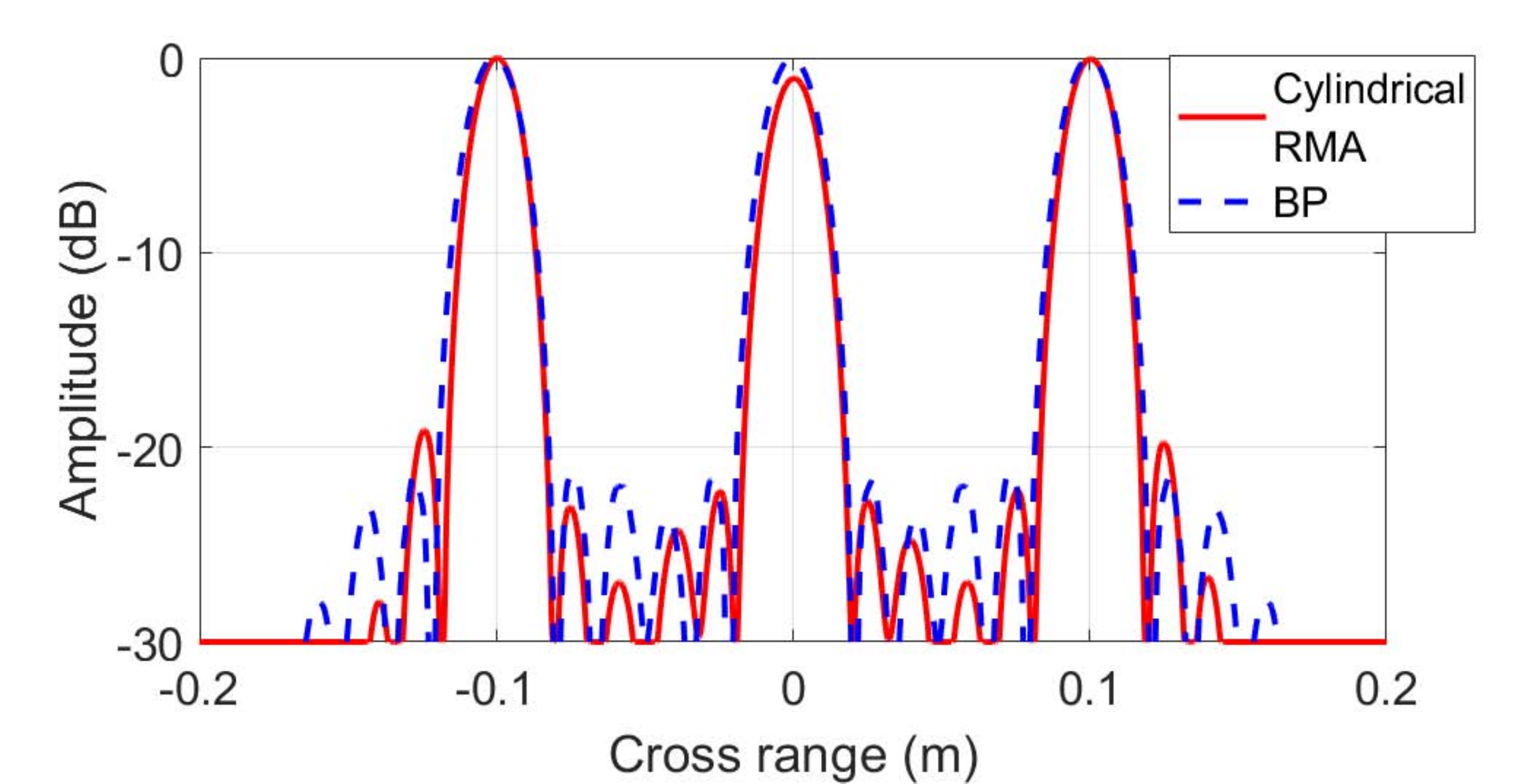}}
	\hfill
	\subfloat[]{\label{c}	
		\includegraphics[width=2.5in]{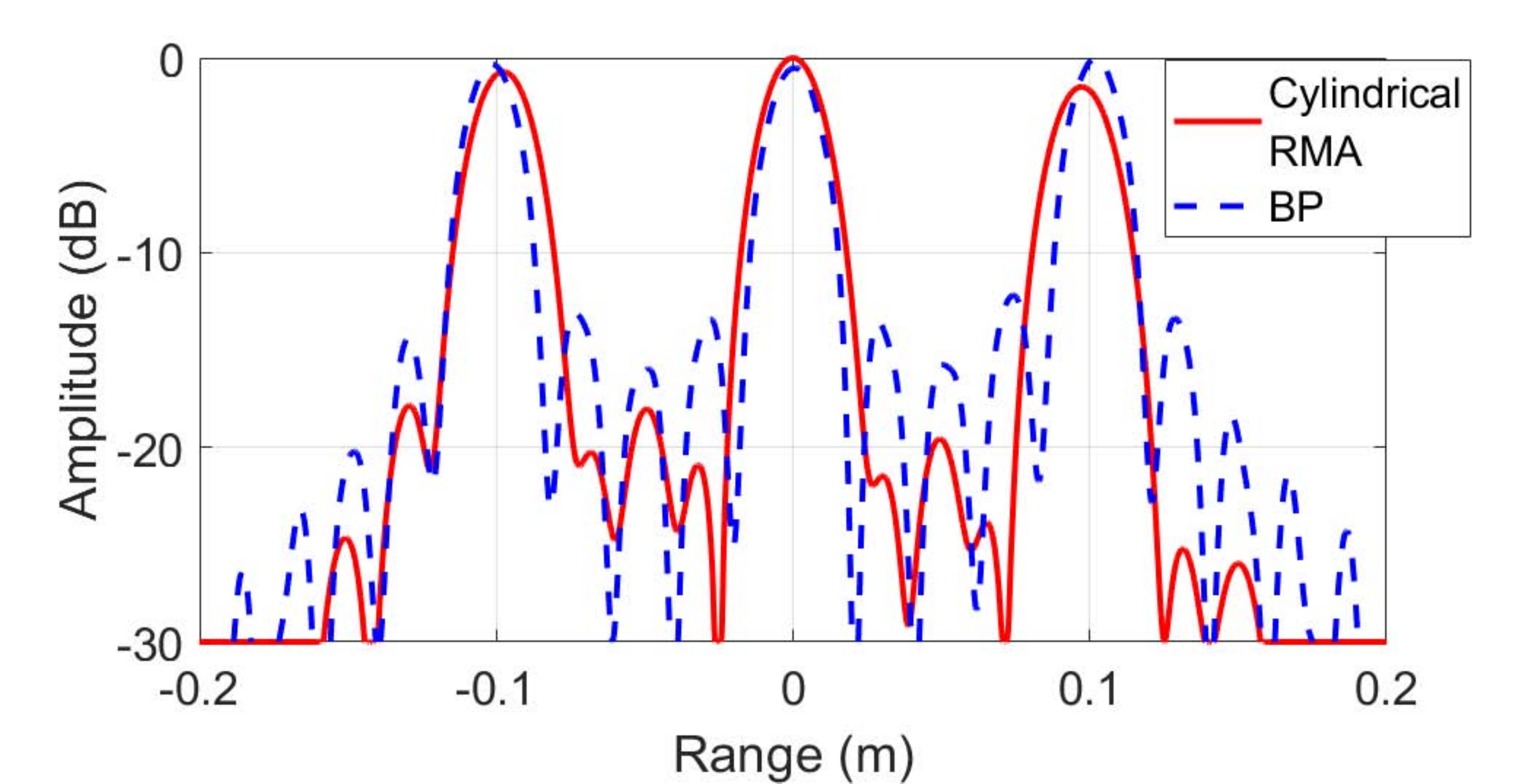}}	
	\\	
	\caption{1-D image corresponding to the (a) height, (b) horizontal, and (c) range dimensions, respectively.}
	\label{1-D slices}
\end{figure}

This section shows the performance of the proposed imaging algorithm using simulations via MATLAB and a full wave simulation tool -- gprMax\cite{gprMax}.
The parameters related to the cylindrical MIMO array  are shown in Table \ref{tab2}.

The 3-D imaging result of the proposed cylindrical RMA is shown in Fig. \ref{3d_mimo}.
The 2-D image slices with respect to the three coordinate planes are given in Fig. \ref{2-D slices}.
To compare the results with BP, the 1-D image slices along the three dimensions are provided in Fig. \ref{1-D slices}.
Note that the image amplitudes of the cylindrical RMA are lower than those of BP which are primarily caused by the truncation of the target spectrum associated with the 2-D interpolations.  
Concerning the resolutions, the two algorithms have  similar performance, as evident in Fig. \ref{1-D slices}.


The computations  of both algorithms are  extensive, with the cylindrical RMA taking about 1,500 seconds, and the BP more than 5,000 seconds, for the above imaging parameters. 
For the cylindrical RMA, most of computation time is taken by the 2-D interpolations in 5-D loops with respect to the antenna array and frequencies. 
This calls for the need to find more efficient approaches in lieu of the 2-D interpolation.

\begin{figure}[!t]
	\centering
	\includegraphics[width=2.5in]{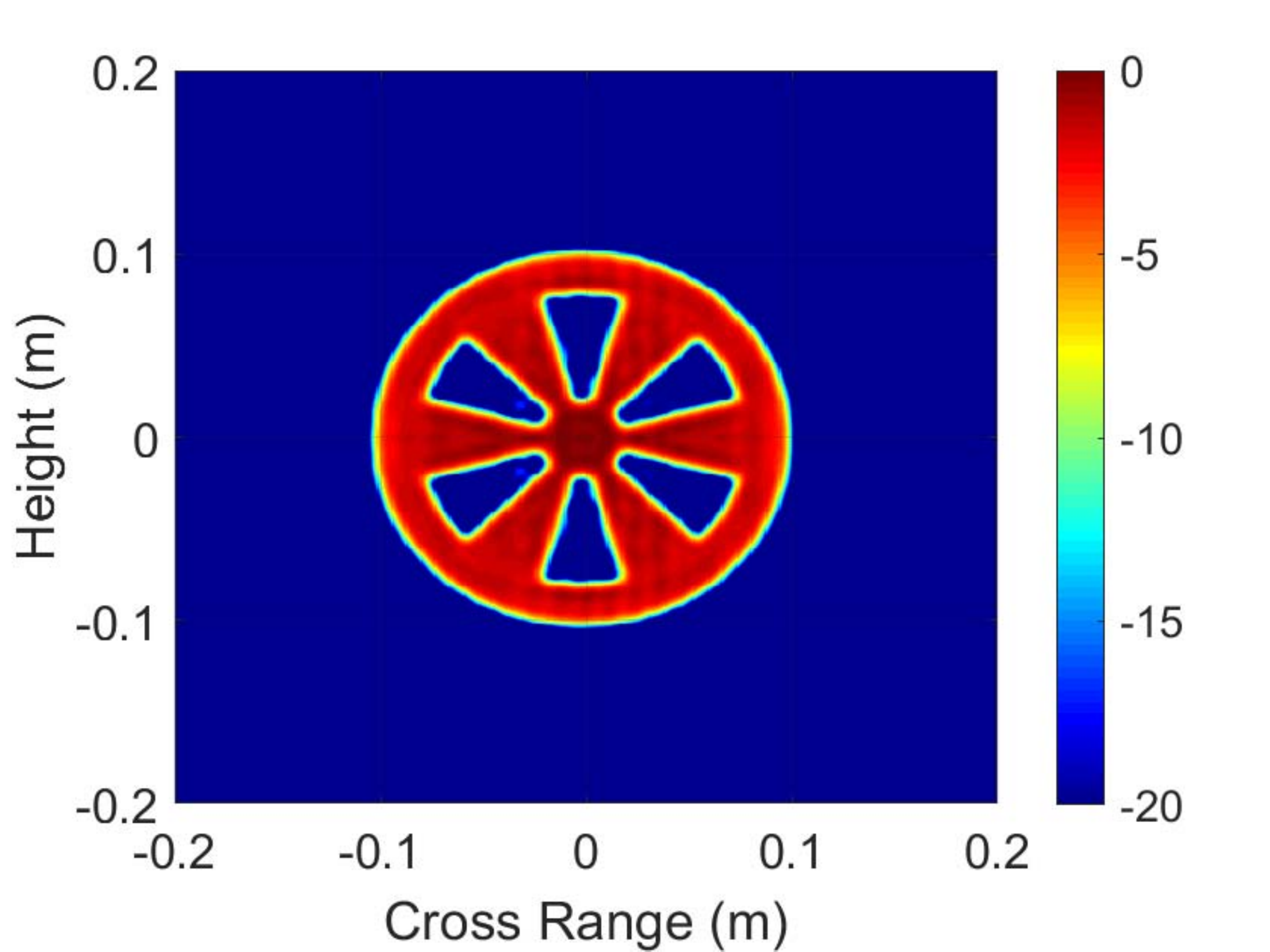}
	\caption{Imaging result of BP.}
	\label{gprmax_bp}
\end{figure}

\begin{figure}[!t]
	\centering
	\includegraphics[width=2.5in]{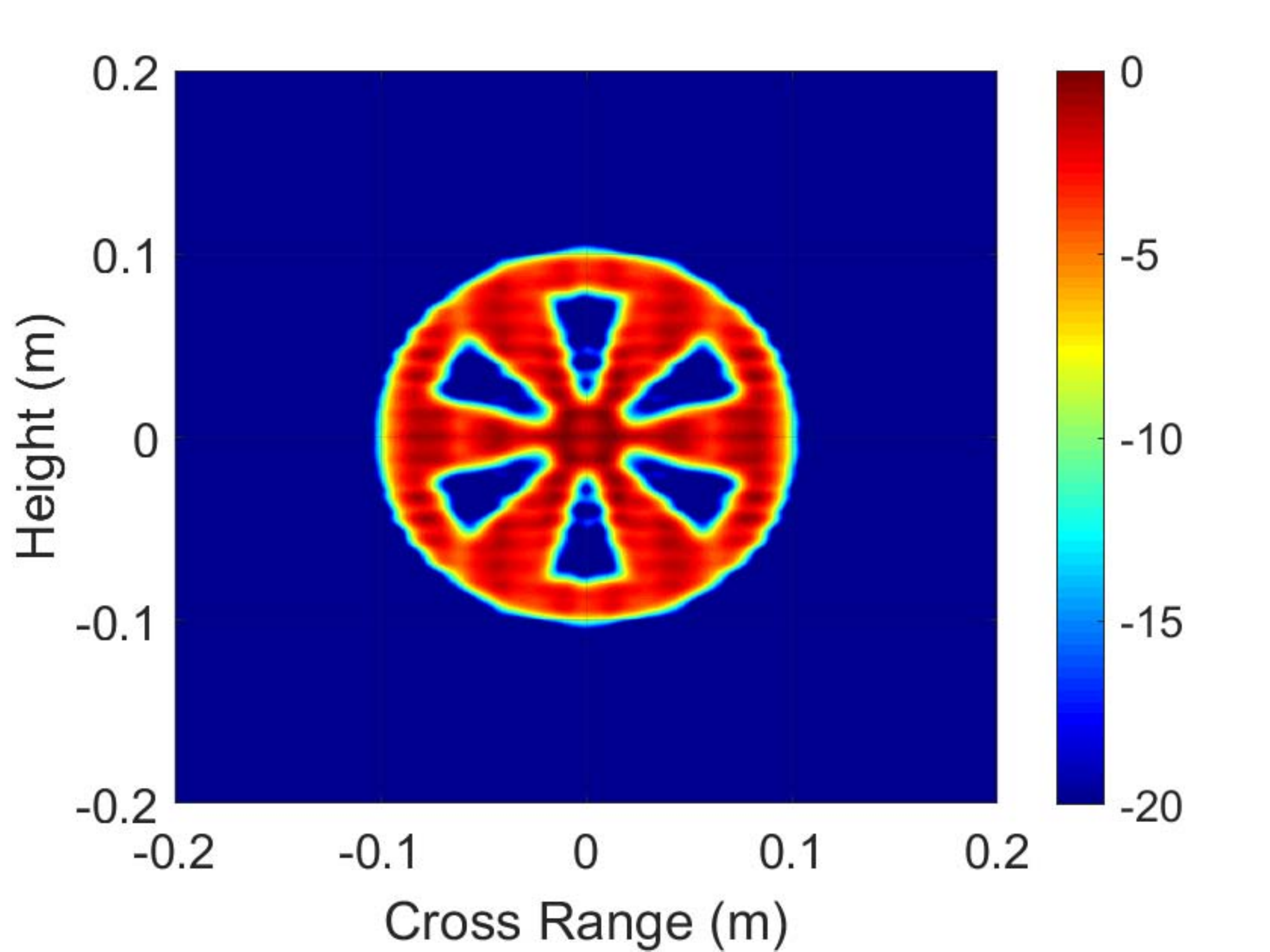}
	\caption{Imaging result of the cylindrical RMA.}
	\label{gprmax_rma}
\end{figure}

Finally, we provide the results using gprMax - a full wave simulation software that solves Maxwell's equations in 3-D using the Finite-Difference Time-Domain (FDTD) method \cite{gprMax}.
To reduce the calculation region in gprMax, we change the radius of the cylindrical aperture to be 0.5m.
The reconstructed images by BP and the cylindrical RMA are shown in Figs. \ref{gprmax_bp} and \ref{gprmax_rma}.
Here, we only present the results with respect to the height and cross range dimensions, which is typically how a 3-D image is shown on a screen, especially for the detection of concealed objects carried by personnel.

Note that the focusing performance of the cylindrical RMA is very close to that of BP, further indicating the effectiveness of the proposed algorithm.


\section{Conclusions}
We proposed a near-field imaging algorithm based on the wavenumber-domain processing, named cylindrical RMA, for a cylindrical MIMO array configuration. 
The spectrum aliasing was examined  via the  zero-filling Fourier transform and  frequency-wavenumber domain  analysis. It was shown that the MIMO  with undersampled subarray can provide desired imaging results under certain sampling and target conditions. 
The requirement for the inter-element spacing of the MIMO array was presented. 
The performance  of the undersampled array is similar to the one obained by a time-domain algorithm, such as BP.
Hence, we can design a MIMO array with either the transmit or the receive array assuming an  undersampled structure. The merit of this design is that we can efficiently utilize  the aperture size to achieve a comparable resolution to a monostatic array. 
Numerical experiments demonstrated the effectiveness of the proposed imaging technique in comparion with the BP algorithm and the conventional wavenumber domain algorithm for a monostatic array with a same aperture geometry.





\ifCLASSOPTIONcaptionsoff
\newpage
\fi

\bibliography{IEEEabrv,full}
\bibliographystyle{IEEEtran}


\end{document}